\documentclass[journal]{vgtc}                

\usepackage{mathptmx}
\usepackage{soul}
\usepackage[normalem]{ulem}
\usepackage{graphicx}
\usepackage{times}
\usepackage{comment}
\usepackage{enumitem}
\usepackage{amsmath}
\usepackage[normalem]{ulem}
\usepackage{xcolor}
\usepackage{caption}
\usepackage{svg}
\usepackage{amsmath}

\usepackage{booktabs}

\usepackage{pifont}
\newcommand{\cmark}{\textcolor{green!80!black}{\ding{51}}}
\newcommand{\xmark}{\textcolor{orange}{\ding{55}}}

\usepackage[bookmarks,backref=true,linkcolor=black]{hyperref} 
\hypersetup{
  pdfauthor = {},
  pdftitle = {},
  pdfsubject = {},
  pdfkeywords = {},
  colorlinks=true,
  linkcolor= black,
  citecolor= black,
  pageanchor=true,
  urlcolor = black,
  plainpages = false,
  linktocpage
}

\onlineid{0}

\vgtccategory{Research}

\vgtcinsertpkg

\title{Thirty-Two Years of IEEE VIS: \\Authors, Fields of Study and Citations}

\author{Hongtao Hao, Yumian Cui, Zhengxiang Wang, and Yea-Seul Kim}
\authorfooter{
\item
 Hongtao Hao is with the University of Wisconsin-Madison. E-mail: hongtaoh@cs.wisc.edu.
\item
 Yumian Cui is with the University of Wisconsin-Madison. E-mail: ycui53@wisc.edu.
\item 
 Zhengxiang Wang is an independent researcher. Email: jackwang196531@gmail.com.
\item
 Yea-Seul Kim is with the University of Wisconsin-Madison. E-mail: yeaseul.kim@cs.wisc.edu.
}

\abstract{The IEEE VIS Conference (VIS) recently rebranded itself as a unified conference and officially positioned itself within the discipline of Data Science. Driven by this movement, we investigated (1) who contributed to VIS, and (2) where VIS stands in the scientific world. We examined the authors and fields of study of 3,240 VIS publications in the past 32 years based on data collected from OpenAlex and IEEE Xplore, among other sources. We also examined the citation flows from referenced papers (i.e., those referenced in VIS) to VIS, and from VIS to citing papers (i.e., those citing VIS). We found that VIS has been becoming increasingly popular and collaborative. The number of publications, of unique authors, and of participating countries have been steadily growing. Both cross-country collaborations, and collaborations between educational and non-educational affiliations, namely ``cross-type collaborations'', are increasing. The dominance of the US is decreasing, and authors from China are now an important part of VIS. In terms of author affiliation types, VIS is increasingly dominated by authors from universities. We found that the topics, inspirations, and influences of VIS research is limited such that (1) VIS, and their referenced and citing papers largely fall into the Computer Science domain, and (2) citations flow mostly between the same set of subfields within Computer Science. Our citation analyses showed that award-winning VIS papers had higher citations. Interactive visualizations, replication data, source code and supplementary material are available at \href{https://32vis.hongtaoh.com/}{\url{https://32vis.hongtaoh.com}}~and~\href{https://osf.io/zkvjm}{\url{https://osf.io/zkvjm}}.
} 


\teaser{
  \centering
  \includegraphics[scale=0.35]{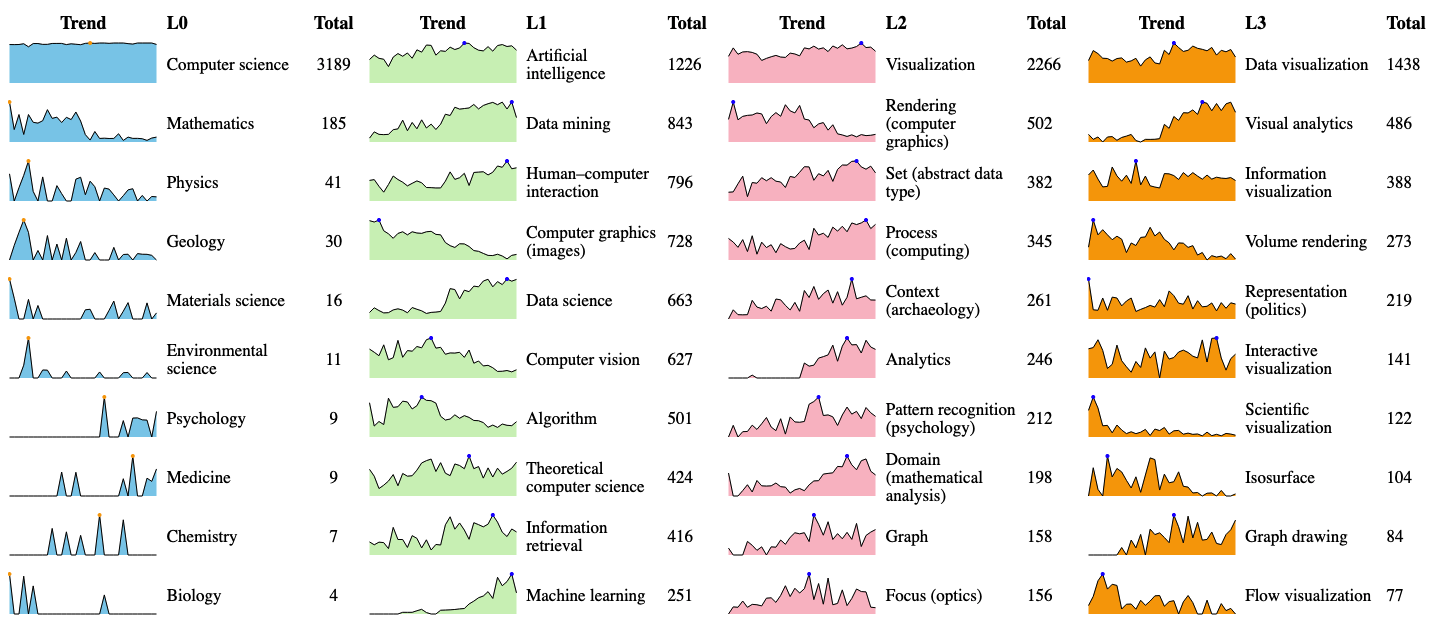}
  \caption{Fields of study at different levels in 3,240 VIS papers across the past 32 years. From L0 to L3, granularity increases. ``Trend'' indicates the proportion of papers falling into a field of study against the total number of papers published in that year. The highest proportion for each field of study is highlighted. ``Total'' indicates the total number of VIS publications falling into a field of study. One paper may contain more than one field of study at the same level, and one field of study may have multiple parent fields. For example, Pattern Recognition belongs to both Computer Science and Psychology.}
  \label{fig:teaser}
}

\definecolor{bias}{HTML}{ff6600}

\definecolor{disp}{HTML}{51A5BA}

\begin{document}



\maketitle
\section{Introduction}

The IEEE VIS Conference (VIS) is the longest-running and the most influential conference in the field of Visualization and Visual Analytics (hereafter collectively called Visualization). It started in 1990, in response to the NSF report of \textit{Visualization in Scientific Computing} \cite{mccormick1988visualization, ieeevis2022}. This marked the beginning of the event as well Visualization as an academic field. The first conference had 52 full papers contributed by 118 unique authors from five countries, namely the US, Germany, Australia, France and Canada. During the past 32 years, VIS has become an international arena: up until 2021, around 6,300 unique authors from 42 countries scattered in all continents across the globe (except for Antarctica) have contributed over 3,200 full papers. Thirty years of history gives us a vantage point to reflect upon the past and think about the future of visualization research.

The visualization community has already started self-reflection. The present unified field of Visualization is a result of several phases of self-transformations. Three sub-conferences, namely Scientific Visualization, Information Visualization, and Visual Analytics, jointly appeared under the umbrella of VISWeek in 2008 and then IEEE VIS in 2013~\cite{chen2021vis30k, isenberg2016vispubdata}. The unification did not stop at the name level; it soon was expanded to the research and organizational level. In 2021, after three years' work, VIS introduced an Area Model where six areas were chosen to represent common research topics in the three subfields and various domain-specific areas \cite{ieeevis2022, acmblog}. This allowed papers to be submitted and reviewed together rather than separately \cite{ieeevis2022}. Visualization is now one unified scientific field at all levels: name, research, and organization. Self-reflection was not only done by the organizing body but also by researchers in the community. Publication metadata \cite{isenberg2016vispubdata}, subject matter \cite{yoshizumi2020review}, figures and tables \cite{chen2021vis30k}, author genders \cite{tovanich2021gender, sarvghad2022scientometric}, publication exploration systems~\cite{tyman2004infovisexplorer, yoon2020conference}, and keywords \cite{isenberg2016visualization} have been the foci of these endeavors. We now have a dataset containing past VIS papers DOIs \cite{isenberg2016vispubdata}, and a repository of all figures and tables in past publications \cite{chen2021vis30k}. We now know that female participation in VIS has been rising \cite{tovanich2021gender, sarvghad2022scientometric}, that geospatial analysis is an important subject matter in recent VIS papers \cite{yoshizumi2020review}, and how keywords in VIS publications evolved and interacted \cite{isenberg2016visualization}. 

There are still, however, some aspects of VIS that we do not know. Perhaps the most fundamental question facing VIS: where does it stand in the landscape of science overall? As of 2022, VIS officially positioned itself within the field of Data Science \cite{ieeevis2022}, and most VIS papers were contributed by Computer Scientists \cite{sarvghad2022scientometric}. However, the remaining questions include: What fields is VIS drawing upon (e.g., Which papers do VIS papers cite?), and where can we see its influences (e.g., Which papers cite VIS papers?)? Apart from the position of VIS in science, we also know little about our authors beyond their genders \cite{tovanich2021gender, sarvghad2022scientometric}. In the introduction to VIS 2022 and 2021, the official conference website mentioned that ``The conference will convene an international community of researchers and practitioners from universities, government, and industry to exchange recent findings ...". The question is, which countries are our authors from, and what is the proportional distribution of different types of author affiliations? Time adds more complexity: are there any temporal changes in answers to the above questions? Addressing these questions has significant implications for VIS as well as for the field of Visualization because we are not able to get a complete picture of who are contributing to VIS, which giants' shoulders does VIS stand on, and which fields VIS is influencing, until we examine authors from different aspects than their genders, and citation flows among referenced (i.e., those referenced in VIS) , VIS and citing (i.e., those citing VIS) papers. 

To address these questions, we collected, merged, cleaned, filled, and aggregated data on VIS authors and publications from various sources. We analyzed country or region (hereafter collectively called country) origins of, affiliations of, and collaborations among authors. We also looked into the fields of study of VIS publications and those of their referenced and citing papers. Based on these data, we built an interactive visualization (\href{https://32vis.hongtaoh.com}{\texttt{https://32vis.hongtaoh.com}}) where viewers can explore temporal trends in fields of study of VIS papers, and the flow of citations based on fields of study. 

Our data show that VIS has been growing steadily in terms of the number of accepted papers, authors, and participating countries. We found that even though the number of participating countries has been increasing, VIS authors are concentrated in a few countries. The dominance of US authors is decreasing, and the number of contributors from China has been rising. We found that the popularity of cross-country collaboration has been constantly increasing. These collaborations, however, are concentrated in a few countries. Although the ratio of collaborations between educational and non-educational affiliations has been growing (with some fluctuations), the ratio of authors from non-educational affiliations has been declining, both within the US and globally. In terms of fields of study, we found that VIS publications mostly concern Computer Science. Similarly, VIS publications were mainly built upon and had their impacts on Computer Science and Mathematical studies. We also found that citations moving in and out of VIS papers mostly flew between a small group of (sub)fields within Computer Science.  

In sum, the contributions of our study are as follows. First, we offer insights into the role VIS is playing in today's scientific landscape and also analyze VIS authors from different perspectives than previous studies \cite{tovanich2021gender, sarvghad2022scientometric}. In addition, our dataset complements those of \cite{isenberg2016vispubdata} and \cite{tovanich2021gender, sarvghad2022scientometric}, helping make more complete data for future scientometric analyses of VIS publications. Lastly, we contribute a workflow through which future researchers may obtain relevant data and conduct similar analyses on other fields. Our results may also inform the general public of the major trends in science development over the past three decades, albeit only in a subdomain of Computer Science.

\section{Background}

Earlier works that treated visualization publications as their subjects of concern were inspired by InfoVis 2004 Contest~\cite{callfor, infovisContest}, which challenged participants to visualize the history of InfoVis. The contest data consisted of 614 InfoVis publications and their over 8,500 references~\cite{infovis2004}. Eighteen submissions from six countries participated. Authors and research areas were the foci of some of the contest papers we found \cite{wong2004spire, ke2004major, keim2004exploring, delest2004exploring, white2004associative}. Similar approaches were applied to visualize VIS authors \cite{WitschardAuthorRanking}, citation motivations in InfoVis papers~\cite{yoon2020conference}, and research topics in TVCG papers~\cite{jiang2016text}.

These works illustrated the importance of publication metadata, which paved the way for scientometric analysis. Another attempt of collecting comprehensive paper metadata in the field of Visualization, i.e., vispubdata.org~\cite{isenberg2016vispubdata}, was completed twelve years later, when VIS was 25 years old. The dataset contained cleaned publication data of VIS papers from 1990 to 2016. Data for 2017-2020 were added later. This work inspired many subsequent scientometric analysis studies \cite{tovanich2021gender, sarvghad2022scientometric, chen2021vis30k} and visualization designs~\cite{zeng2021vistory, wang2019vispubcompas}. Specifically, based on \cite{isenberg2016vispubdata}, scholars collected tables and figures in past VIS publications~\cite{chen2021vis30k}, and analyzed VIS author genders and collaborations~\cite{sarvghad2022scientometric, tovanich2021gender}. They found that female participation in VIS had been constantly rising but gender gaps remained. For example, female authors were less likely to be the last authors \cite{tovanich2021gender}, and gender balance at VIS was predicted to be achieved only half a century later \cite{sarvghad2022scientometric}. 

The work by Sarvghad et al. \cite{sarvghad2022scientometric} looked very similar to ours as we both examined VIS authors and research areas. A closer examination revealed that we were different. \cite{sarvghad2022scientometric} analyzed collaborations from the perspectives of authors' research areas, genders, and institutions, whereas we focused on collaborations based on author country origins and the types of their affiliations. Besides, \cite{sarvghad2022scientometric} inferred fields of study from authors' affiliations whereas our fields of study classification was directly derived from each paper itself. That being said, our study corroborated some major findings in \cite{sarvghad2022scientometric} but from different angles.

Apart from authors, keywords were an important element in publications. Co-word analysis was applied to keywords in VIS publications from 1990 to 2015 \cite{isenberg2016visualization}. Key themes among VIS papers, the relationships among these themes, and how keywords emerged and evolved were examined. It was found that pre-defined keywords provided on the Precision Conference System had mainstream topics, whereas topics extracted from keywords in original VIS PDFs did not. Similar work was conducted earlier for CHI publications \cite{liu2014chi}. 

Some scientometric studies on visualization and related fields, for example, InfoVis~\cite{henry200720}, CSCW~\cite{correia2018scientometric}, CHI~\cite{bartneck2009scientometric} and IndiaHCI~\cite{gupta2015five}, instead of focusing on one or two aspects, did a comprehensive overview of their fields of interest.

\section{Data collection and processing}

In this section, we detail how we collected and processed data on VIS papers. All data collection was completed in early June of 2022. 

\begin{figure*}[!htbp]
\centering
\includegraphics[width=\linewidth]{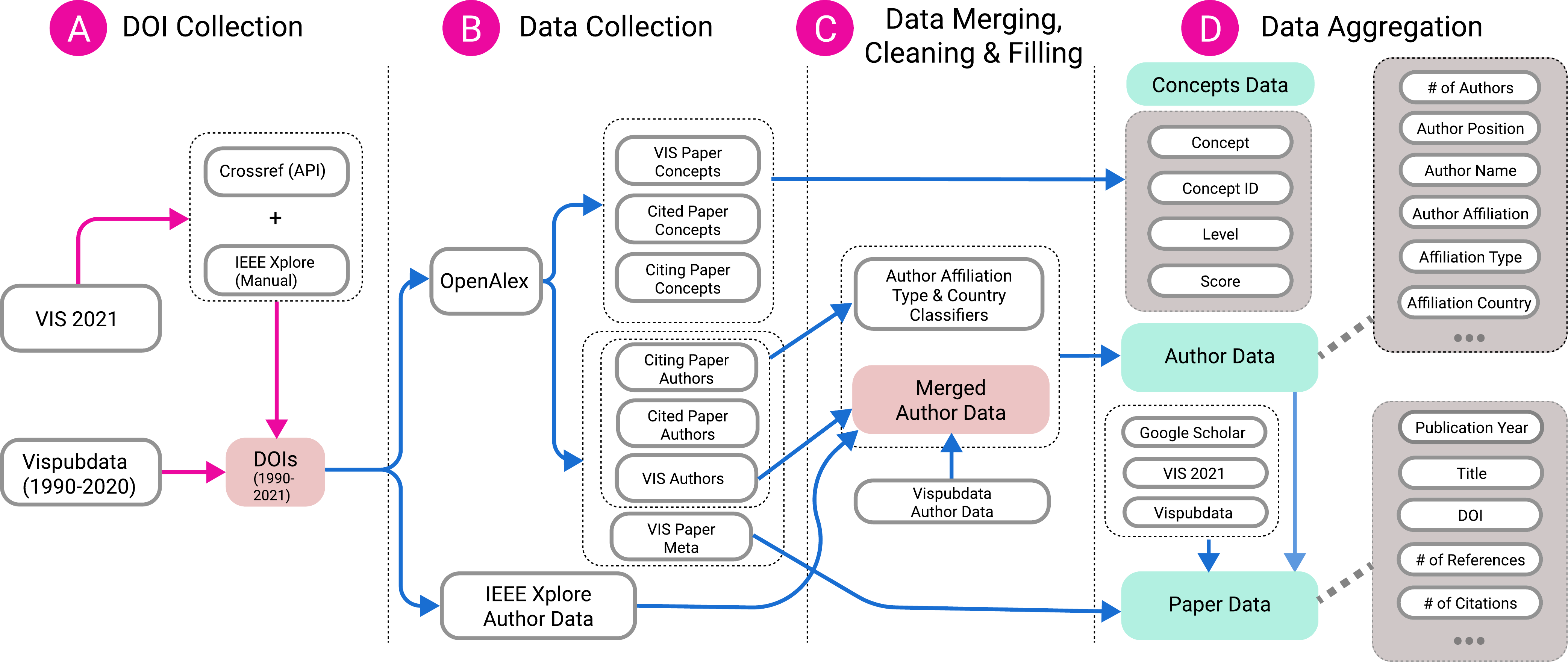}
\caption{Data processing pipeline. We started from Vispubdata and VIS2021 to get the paper DOIs, with which we obtained data on relevant papers from OpenAlex and IEEEXplore. We then merged and cleaned data and filled in missing data. Some of the paper data, for example, whether a paper is a cross-country or a cross-type collaborative paper, came from author data. The final outputs were three major data files: authors, concepts, and paper meta.}
\label{fig:pipeline}
\vspace{-1.5em}
\end{figure*}

\subsection{DOI collection}

Our study analyzed full papers of VIS published between 1990 and 2021. In the following, we describe how we identified their DOIs.

We obtained from vispubdata.org \cite{isenberg2016vispubdata} the DOIs and titles of 3,394 VIS papers published from 1990 to 2020. Following the practice of \cite{chen2021vis30k}, we only included conference and journal publications and case study papers whereas excluded posters, panels, and keynote documents (i.e., those with the paper type of ``M'' as coded in \cite{isenberg2016vispubdata}). We excluded these publications because (1) neither IEEEXplore~\cite{ieeexplore}, a digital library containing papers published by IEEE and its partners, nor Vispubdata~\cite{isenberg2016vispubdata} had complete data on them. For example, both sources lacked posters from 2017 and 2019; and (2) they differ qualitatively from full papers. After this exclusion, we had 3,073 papers. We found no duplicates in their titles but found one paper with an invalid DOI (10.0000/00000001) which we removed from further analysis. The final result of this step was 3,072 paper DOIs from Vispubdata~\cite{isenberg2016vispubdata}.

170 full papers presented at VIS2021 were not part of  \cite{isenberg2016vispubdata}. We collected their titles from the official website of VIS2021. We found no duplicates in these titles. Using the package of habanero~\cite{habanero}, we obtained these papers' DOIs from Crossref~\cite{crossref} based on title queries backed up by manual data collection from IEEEXplore if automation failed (See \textit{Supplementary Material Methods 1.1} for details.). We inspected the DOIs of these 170 VIS2021 papers and found all of them were valid. Thus far, we had 3,242 (3,072 + 170) VIS papers. 

\subsection{Data collection}

After identifying the papers to study, the next step was to collect relevant data on them. Given the motivation behind our study, we needed information on VIS papers from these aspects: (1) VIS authors, and (2) fields of study of VIS, referenced, and citing papers. For authors, we needed author names, positions, and affiliations. For referenced and citing papers, we wanted numbers of counts, paper titles, and fields of study. Neither IEEEXplore nor Vispubdata \cite{isenberg2016vispubdata} had complete information on all these variables, so we collected the data from OpenAlex~\cite{priem2022openalex} as they offered much richer data on author affiliations and fields of study. Other outlets such as PubMed, Web of Science, Scopus, Crossref, and Google Scholar did not offer the metadata we wished to collect. For detailed accounts on how we ended up choosing OpenAlex, refer to \textit{Supplementary Material Methods 1.2}. Table~\ref{tab:features} shows availability of key features on different academic databases. 

Among all 3,242 papers, we were able to identify 3,240 of them in OpenAlex through a combination of title and DOI queries, whereas 2 papers, namely 10.1109/VISUAL.1990.146412 and 10.1109/VISUAL.2003.1250379, did not exist in the OpenAlex database. We excluded these two publications from our following analyses. In sum, our collection consisted of 3,240 VIS full papers published from 1990 to 2021, including data on their authors, fields of study, and paper metadata. Examples of variables we had are shown in Fig.~\ref{fig:pipeline}.

\begin{table}
\caption{\label{tab:features}Availability of key features on popular scholarly databases. ``Fields'' stands for Fields of Study.}
\small
\centering
\begin{tabular}{ccccccccc}
\toprule
 & Free & Author & Cit. & Ref. & Fields & Maintained & API\\
\midrule
WoS & \xmark & \cmark & \cmark & \cmark & \cmark & \cmark & \cmark\\
Scopus & \xmark & \cmark & \cmark & \cmark & \cmark & \cmark & \cmark\\
PubMed & \cmark & \xmark & \cmark & \xmark & \xmark & \cmark & \cmark\\
JSTOR & \cmark & \xmark & \xmark & \xmark & \xmark & \cmark & \cmark\\
Crossref & \cmark & \xmark & \cmark & \cmark & \xmark & \cmark & \cmark\\
Google & \cmark & \xmark & \cmark & \xmark & \xmark & \cmark & \xmark\\
Semantic & \cmark & \xmark & \cmark & \cmark & \xmark & \cmark & \cmark\\
MAG & \cmark & \cmark & \cmark & \cmark & \cmark & \xmark & \cmark\\
OpenAlex & \cmark & \cmark & \cmark & \cmark & \cmark & \cmark & \cmark\\
\bottomrule
\end{tabular}
\vspace{-2em}
\end{table}

\subsection{Data merging, cleaning, and filling}

For author data, we relied on IEEEXplore and OpenAlex, coupled with Vispubdata \cite{isenberg2016vispubdata} for cross-validation. From IEEEXplore, we collected author information, including the number of authors, author position, author name, IEEE author ID, and author affiliations. From OpenAlex, we collected the same data (except for IEEE author ID), plus author affiliation type and affiliation's alpha-2 (ISO 3166) country code (e.g., US, CN, DE, etc). 
We compared the number of authors for each paper in the two datasets and found IEEEXplore was incorrect in one paper (DOI: 10.1109/TVCG.2008.157): it listed five authors, but there were only four when we checked the PDF. Also, IEEEXplore did not contain information about the paper of 10.1109/VIS.1999.10000. We fixed these errors and updated IEEEXplore author data. Among 12,423 authors in the dataset obtained from IEEEXplore, only 337 (2.7\%) contained more than one affiliation. For consistency and simplicity, we only included their first affiliation in our analyses.

We merged the IEEEXplore and OpenAlex author data based on exact matching of DOI and fuzzy matching of author names. We merged the two datasets for two reasons: 1) IEEEXplore missed affiliation information for 167 authors, some of which could be found in OpenAlex dataset; and 2) We could cross-validate author affiliations in both datasets if necessary. We then compared the number of authors for each paper in our merged dataset with the DBLP~\cite{dblp} author information collected by~\cite{isenberg2016vispubdata}. We identified 17 instances where the two datasets disagreed. We checked the original PDFs of these 17 publications and found our merged author data were incorrect in 4 papers and DBLP was incorrect in 13 papers. We updated our merged dataset with correct author data. The final merged author data consisted of 12,428 authors. Note that ``authors'', unless we specify they are ``unique authors'', might be duplicates. For example, if one author is present in 10 VIS publications, we consider them as 10 authors. This is because we were more interested in author country origins and affiliations than authors per se. Deduplicating authors did not make sense in our study because authors may change their affiliations.

The IEEEXplore portion of the merged author dataset missed affiliation information for 181 authors. We filled in these missing data based on author descriptions and email addresses in the original paper PDFs, author profiles on IEEEXplore, and open web search. For papers where we were uncertain of our conclusions, we requested via email from original authors information regarding their affiliations at the time of publication. When we were manually collecting affiliation data, we filled in the author affiliation type and country origin following OpenAlex's criteria. We also corrected the errors we noticed in author names, author affiliations, and affiliation country codes. After this procedure, 66 authors still missed affiliation information. OpenAlex had this information, which we manually validated and found two errors in author affiliation and one error in author name. After correcting these errors, we used OpenAlex's data to fill in the missing affiliation data. Up until this stage, we made sure all 12,428 authors had complete affiliation information. 

These merged author data, however, were incomplete in author affiliation type and affiliation country code. Among 12,428 authors, 2,498 (20.1\%) missed affiliation type and 2,328 (18.7\%) lacked country information. There were also problems for observations that were complete in these two variables: the available data from OpenAlex on affiliation type and country codes were based on affiliation strings provided by OpenAlex. These strings, however, were slightly different from those on IEEEXplore. Therefore, even for rows where affiliation type and country data were complete, we were not 100\% sure that they were correct information for the actual authors. Fortunately, IEEEXplore provided affiliation information for 99\% of all authors; the rest were added through our above mentioned data merging and filling procedure. Since IEEEXplore is the official data source on VIS authors, we regarded their data as reliable. 

To automatically infer affiliation type and country code from affiliation names offered by IEEEXplore, we utilized classifiers. We assumed that OpenAlex's classifications of affiliation type and country codes based on affiliation strings were mostly accurate, and they were: we randomly selected 100 observations where OpenAlex data were complete and we concluded that the mappings were 99\% correct for affiliation types and 98\% correct for country codes. We built two separate multiclass text classifiers with logistic regression~\cite{scikit-learn} based on author data in 3,240 VIS papers, 39,817 unique referenced papers, and 60,272 unique citing papers. After deduplication, we obtained 73,199 feature-label pairs for affiliation type classification and 75,706 pairs for country code classification. We randomly split the data into the training set (80\%) and the test set (20\%). The affiliation type classifier reached a test set accuracy of 92.4\% (precision, recall and F1 scores were almost the same). This accuracy increased to 95.0\% if we only considered binary classification, i.e., education versus non-education. The country code classifier reached a test set accuracy of 93.1\%, which was almost the same for precision, recall, and F1 scores. We applied these two classifiers to our author affiliation data obtained from IEEEXplore. After predictions were completed, we randomly selected 100 rows and checked the prediction accuracy. 99 out of 100 predicted country codes were correct. The prediction for affiliation type was 95\% correct. If we only considered a binary classification, the accuracy was 99\%. The reason why country code predictions were nearly perfect was that many affiliation names on IEEEXplore already contained country information. 

\subsection{Data aggregation}

\begin{figure*}[!t]
\centering
  \includegraphics[width=\linewidth]{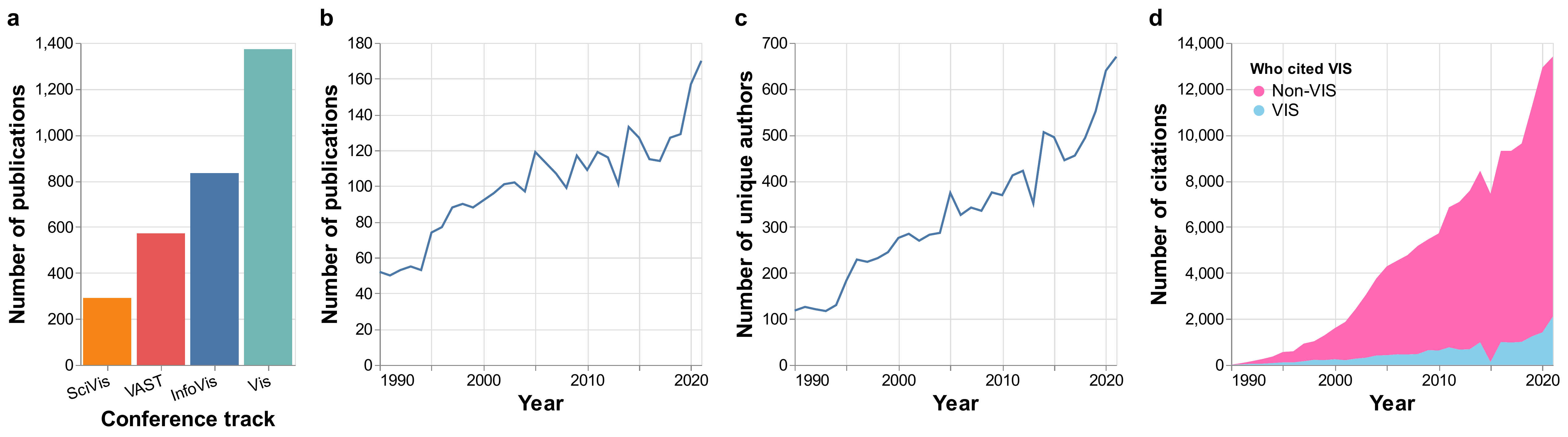}
  \includegraphics[width=\linewidth]{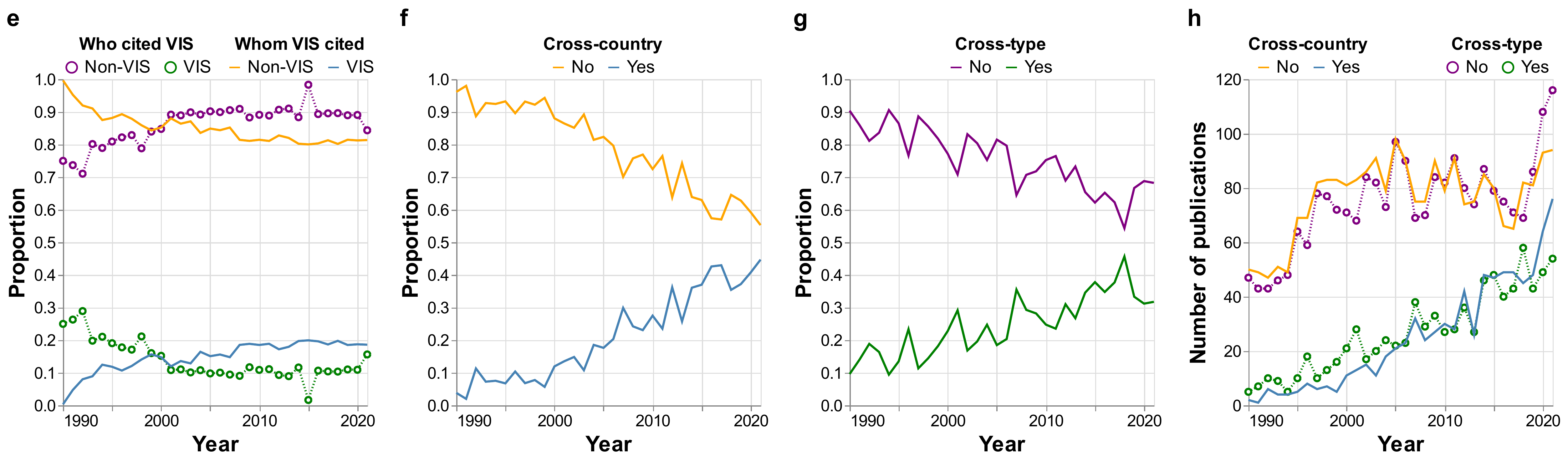}
    \caption{General trends. \textbf{a} Number of publications of each of the five tracks. \textbf{b} Number of VIS publications each year. \textbf{c} Number of participating unique authors each year. Note that among 12,428 authors, 110 of them did not have OpenAlex author ID so they were not included here. \textbf{d} Number of citations from VIS and non-VIS papers. \textbf{e} Proportion of VIS and non-VIS papers that were referenced in, and citing, VIS publications. \textbf{Note that in d. and e., the data is year-by-year rather than cumulative.} \textbf{f} and \textbf{g} Proportion of cross-country collaboration, and cross-type collaboration. \textbf{h} Number of publications from cross-country collaboration and cross-type collaboration.}
\vspace{-2mm}
\label{fig:trends}
\end{figure*}

From Vispubdata~\cite{isenberg2016vispubdata}, we collected for each paper its title, DOI, and publication year. We also obtained conference track information, i.e, InfoVis, SciVis, VAST, and Vis. Distribution of publication counts among conference tracks before VIS2021 can be found in Fig.~\ref{fig:trends} (a). Note that this figure does not include VIS2021 publications because starting from 2021, VIS no longer distinguishes between these tracks. We used ``VIS'' as an umbrella term to indicate publications in all these four tracks and also the VIS2021 papers. 

From OpenAlex, we collected each paper's number of references and citations. Based on title queries backed up by DOI queries, we collected citation counts on Google Scholar, which were used (1) as validation of citation data from OpenAlex and (2) in our citation analyses. We were able to identify all 3,240 papers. From the official website of VIS2022~\cite{vis2022awards}, we obtained the historical information on award-winning papers. We considered a paper as an award-winning one if it received Best Paper Award, Honorable Mention Award, or Best Case Study Award. We excluded Test of Time Awards because they are released more than ten years after the publication of the awarded paper. Based on our author data, we decided for each paper, whether it is (1) cross-type collaboration, i.e., collaborations between educational affiliations and non-education affiliations, (2) cross-country collaboration, and (3) involving authors from the United States. These variables were important in our analyses of collaboration patterns in VIS and also the changing role of US authors.  

To answer the question of ``where VIS stands in science'', we need to know (1) which topics VIS publications cover, (2) on which fields VIS is built, and (3) which fields VIS is impacting. OpenAlex's concepts data included around 65,000 fields of study at various levels and therefore are optimal for the present study. OpenAlex represented fields of study as concepts. Concepts are hierarchical such that they belong to different levels. Level 0 concepts such as Computer Science, Mathematics, and Psychology do not have ancestors. L1-L3 represent more granular fields; the larger the number, the more granularity. Note that one concept may have multiple parents. For example, the L2 concept of Visualization belongs to two L0 concepts, i.g., Computer Science and Engineering, and to three L1 concepts, i.g., AI, Data Mining, and Mechanical Engineering. Also note that concept tagging is level-independent. Therefore, for example, if a paper is tagged with the L2 concept of Visualization, it does not necessarily mean that this paper has all (or even any) of Visualization's parent concepts at L1. Detailed statistics of these concepts are available in Table~\ref{tab:concepts}. \textit{Supplementary Material Methods 1.3} has more information on OpenAlex concepts tagging procedure. 

Based on Fields of Study data from Microsoft Academic Graph, OpenAlex trained a classifier that assigned concepts of different levels to a paper~\cite{concept-tagger}. Assigned concepts had associated scores in a way that concepts of a higher score were a better representation of a paper. Concepts with a score lower than 3.0 were not assigned. We collected concepts data from OpenAlex for 3,240 VIS papers, 39,817 unique referenced papers, and 60,272 unique citing papers.
\section{Results}

We present major findings in this section. Figures were produced with Altair~\cite{vanderplas2018altair}, D3.js~\cite{bostock2011d3}, and Seaborn~\cite{waskom2021seaborn}.

\begin{figure}[!t]
\centering
\vspace{-2em}
  \includegraphics[scale=0.3]{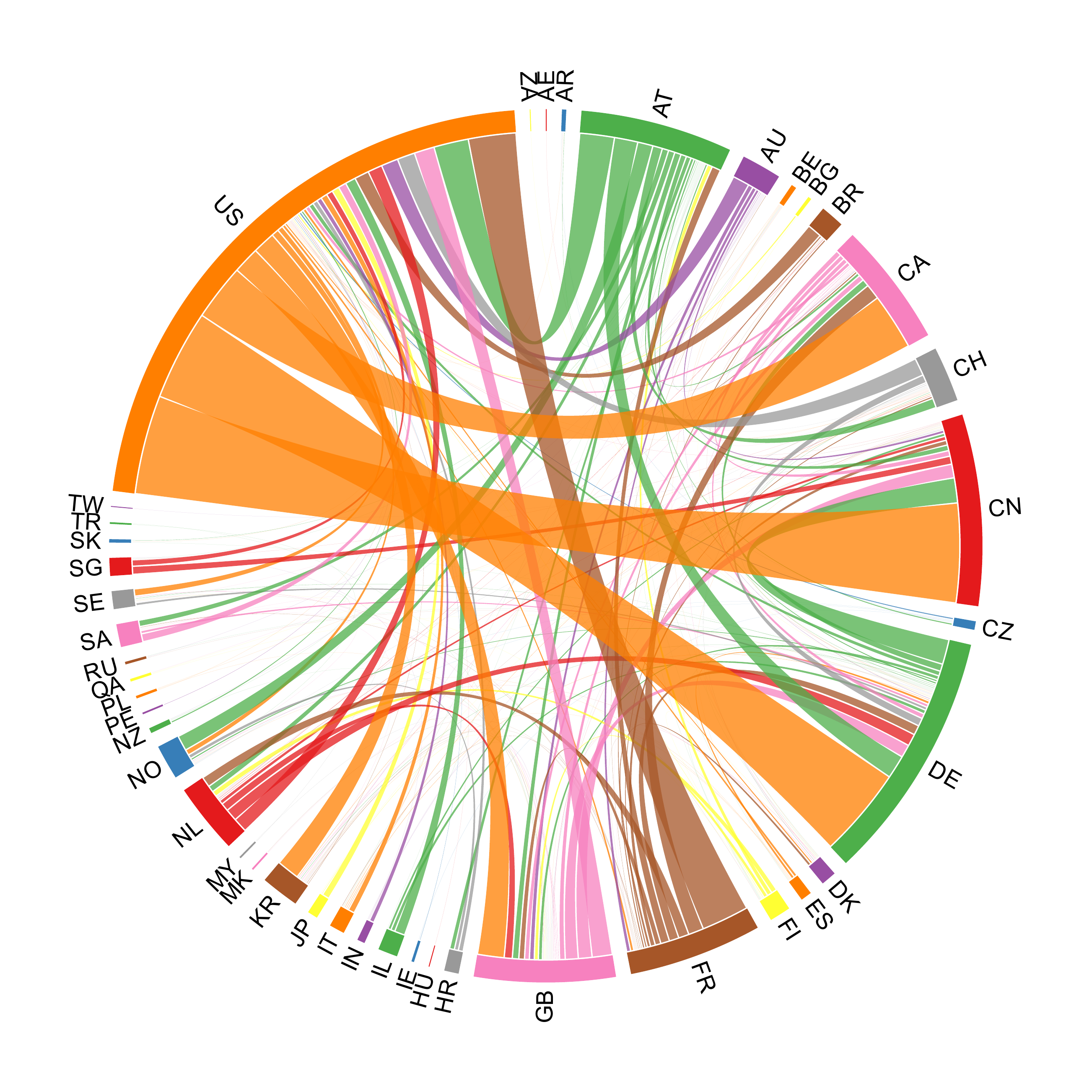}
  \vspace{-1em}
  \caption{VIS author collaboration network. Each arc length indicates the number of times a country or region appeared in a collaboration chord. For example, if a paper involves five authors (three from US, one from FR, and one from CN), then there will be three chords: US-FR, US-CN, and FR-CN. The color of each chord is assigned randomly. The top ten most collaborative countries are: US, Germany (DE), China (CN), Austria (AT), UK (GB), France (FR), Canada (CA), Netherlands (NL), Switzerland (CH), and Australia (AU). These countries were present in 98\% of all the collaboration chords, and 71\% of all collaborations throughout the history of VIS occurred among these countries.}
\vspace{-2em}
\label{fig:author_network}
\end{figure}


\subsection{General Trends}

\begin{figure*}[!t]
\centering
\includegraphics[width=\linewidth]{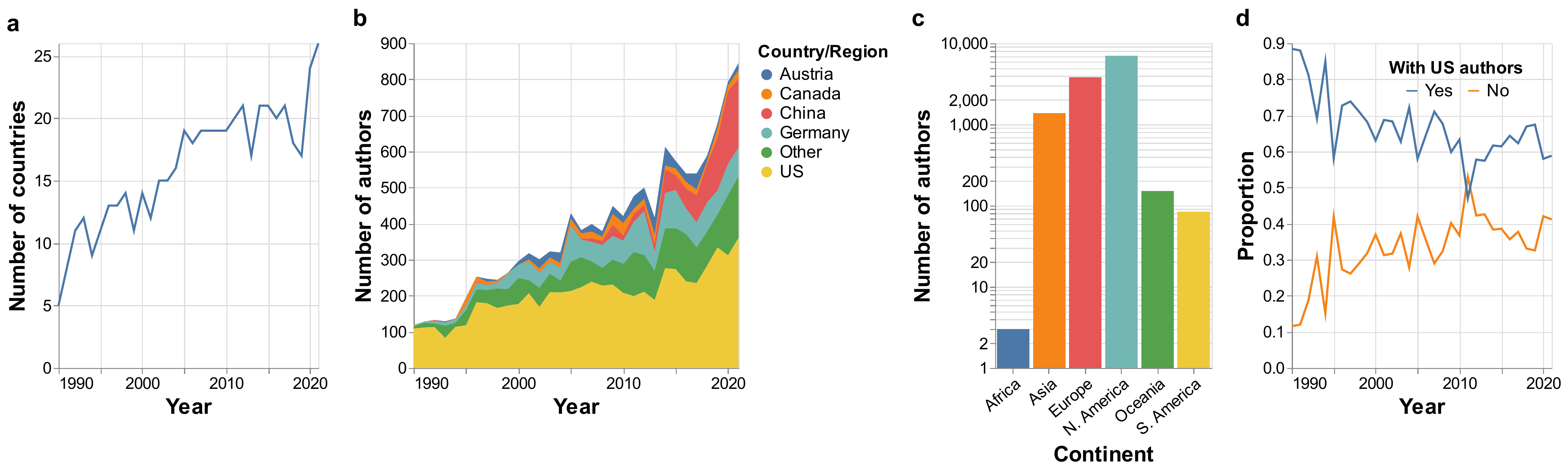}
\includegraphics[width=\linewidth]{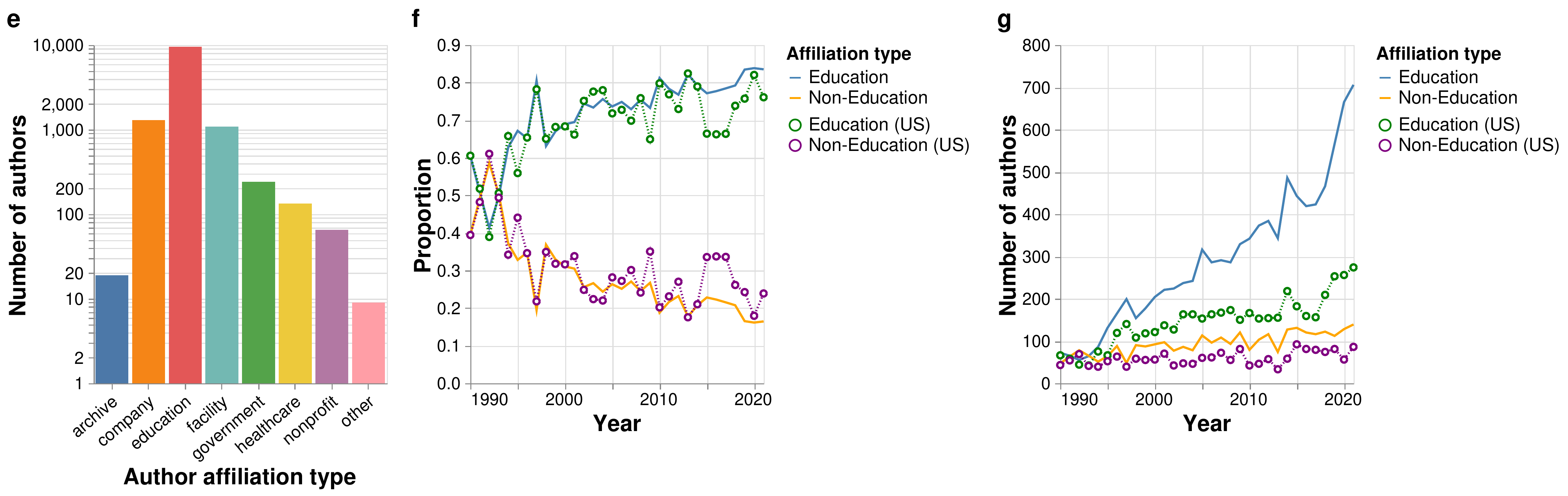}
\caption{Authors. \textbf{a} Number of participating countries by year. \textbf{b} Temporal changes in the number of authors from top five countries. Note that the Y-axis shows yearly number of authors, not cumulative counts. \textbf{c} Distribution of authors by continent. \textbf{d} Temporal changes in the proportion of papers involving US authors. \textbf{e} Distribution of author affiliation types. \textbf{f} and \textbf{g} Proportion and absolute numbers of authors from educational and non-educational affiliations; Both the global (including the US) the US data alone were plotted.}
\vspace{-2mm}
\label{fig:author}
\end{figure*}

We found three trends: VIS is becoming more popular, impactful, and collaborative. 

The rise in popularity is evident in the increasing number of publications and unique authors. As Fig.~\ref{fig:trends} (b) shows, in the first conference in 1990, there were 52 full papers. This figure grew to 170 in 2021. With an acceptance rate of around 25\% in all sub-conferences in the past decade \cite{isenberg2016vispubdata}, this growth indicates that there have been an increasing number of submissions to VIS. There was also an increase in the number of unique authors as shown in Fig.~\ref{fig:trends} (c). In 1990, there were 118 unique authors, whereas this number grew to 670 in 2021, a 468\% jump. The growth of VIS popularity was also evident in the increasingly diversified country origins of authors, which we discuss later.

As evident in Fig.~\ref{fig:trends} (d) and (e), the impact of VIS can be seen in the increase in both the absolute number and proportion of citations attributed to non-VIS publications. In 2021, VIS papers were cited 11,309 times by publications outside of the VIS community, almost doubling the number in 2011 (6,091). The proportion of citations attributed to non-VIS papers also showed an upward trend. In the first five years, this proportion fluctuated between 70\% and 80\%; since 1999, it had been above 80\% and even above 90\%. In 2021, it was 84\%. The proportion of non-VIS papers in referenced papers, however, showed a slightly downward trend, indicating a growing trend that VIS had been built upon past work in the community itself. 

The collaborative nature of VIS is evident in the increase in (1) the average number of authors, and (2) the proportion of papers resulting from cross-country and cross-type collaborations.

In the first ten years, VIS publications were contributed by on average two to three authors. In the next decade, i.e., from 2000 to 2010, this number was between three and four. Starting from 2012, the average number of authors per paper had been constantly above four, peaking at 5.3 in 2019. In 2021, this number was around 5. (See \textit{Supplementary Material Results 2.1} for details.)

In the first conference in 1990, among 52 full papers, only 2 were by authors from different countries as indicated by Fig.~\ref{fig:trends} (h): one was a US-Germany collaboration and the other was between the US and Canada. As demonstrated in Fig.~\ref{fig:trends} (f), in the first half of VIS history (1990-2005), the yearly proportion of cross-country collaboration papers was always below 20\%. After 2006, this figure remained above 20\%; since 2014, above 30\%. The most recent year, i.e., 2021, saw a historical high: 45\% of all papers were cross-country collaborations. These collaborations, however, were highly concentrated, as can be seen in Fig.~\ref{fig:author_network}. After deduplicating the collaboration pairs in each paper such that if a paper had five authors (e.g., four from the US and one from China), we assigned only one pair, namely ``US-CN'' to it, we had 1,218 collaboration pairs with 2,436 nodes (i.e., collaborating countries). The top 10 most active countries in cross-country collaboration, namely, the US, Germany, China, Austria, the UK, France, Canada, Netherlands, Switzerland, and Australia, appeared in 1197 (98\%) pairs among all 1,218 pairs. Together, collaborations among these most collaborative countries were responsible for 71\% of all collaborations throughout the history of VIS. 

Similar growth was observed in cross-type collaborations. As shown in Fig.~\ref{fig:trends} (g), the proportion of cross-type collaboration fluctuated between 10\% and 30\% in the first half of VIS (1990-2006). In the years following 2007, this proportion was always above 20\%, peaking at 46\% in 2018. Since then, the proportion has dropped a little bit. In 2021, 32\% of all publications were cross-type collaborations. 

\subsection{Country origin and types of author affiliations}

In 1990, 118 unique authors from only five countries (Fig.~\ref{fig:author} (a)) participated in VIS; 108 (91.5\%) of them were from the US (Fig.~\ref{fig:author} (b)), with the remaining ten authors' country origins scattered in Germany (4), Australia (3), France (2) and Canada (1). Beginning from the third conference in 1995, the number of participating countries had always been above ten, and since 2003, at least fifteen. Since 2008, this number fluctuated around twenty, peaking in VIS 2021 where there were 26 participating countries. 

Although authors from diverse places participated, the majority of them came from only a few countries. The top five sources of authors, namely the US (53\%), Germany (13\%), China (8.2\%), Austria (4.2\%), and Canada (3.5\%) were responsible for 82.1\% of all 12,428 authors. If we consider the top ten, which would add the UK, France, Netherlands, Switzerland, and Australia, this figure jumped to 93\%. In terms of continents, almost all authors (98\%) came from North America (56.4\%), Europe (30.6\%) and Asia (11.1\%) as depicted by Fig.~\ref{fig:author} (c). Throughout the past 32 years, only 3 authors came from Africa, 84 (0.7\%) from South America, and 151 (1.2\%) from Oceania. 

There were some redistributions in country origins among top author sources. As shown in Fig.~\ref{fig:author} (b), the dominance of the US has been declining, and the number of authors from China has been increasing. In terms of the cumulative number of authors, China is the third-largest source, only after the US and Germany. If we look at the number of authors in yearly terms, China overtook Germany in 2017 as the second-largest author source and has remained in that place since then. Taking into consideration the 258 authors from Hong Kong SAR separately did not change the overall statistics: Mainland China is still the third-largest author source in terms of the cumulative count and has been the second-largest in terms of yearly counts since 2019.

Looking at the US involvement in VIS from another angle, however, revealed that the US dominance still existed. As shown in Fig.~\ref{fig:author} (d), in the first five conferences, i.e., between 1990 and 1994, the proportion of papers involving at least one US author had always been above 80\%, except for in 1993 when the proportion plunged to 69\%. From 1995 to 2010, this number fluctuated between 60\% and 70\% and then dropped to a historical low at 47.1\% in 2011. The number then went back again and has been between 50\% and 70\% ever since. In 2021, 58.8\% of all accepted papers involved at least one US author. In terms of the cumulative count in the past 32 years, 64.9\% of all 3,240 VIS publications had US involvement. 
\begin{figure*}[!h]
\centering
  \includegraphics[scale = 0.55]{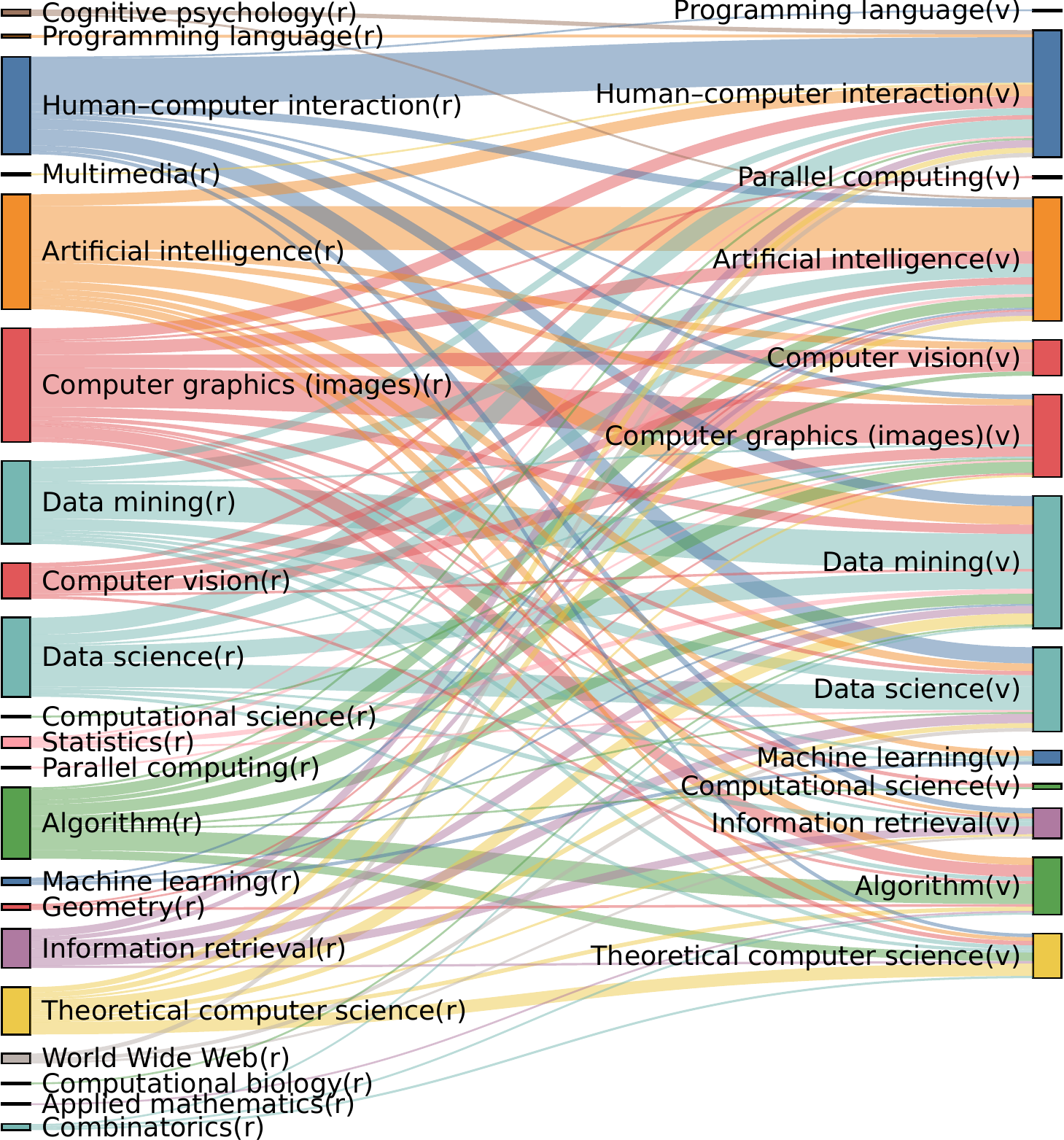}
  \hspace{2em}
  \includegraphics[scale = 0.55]{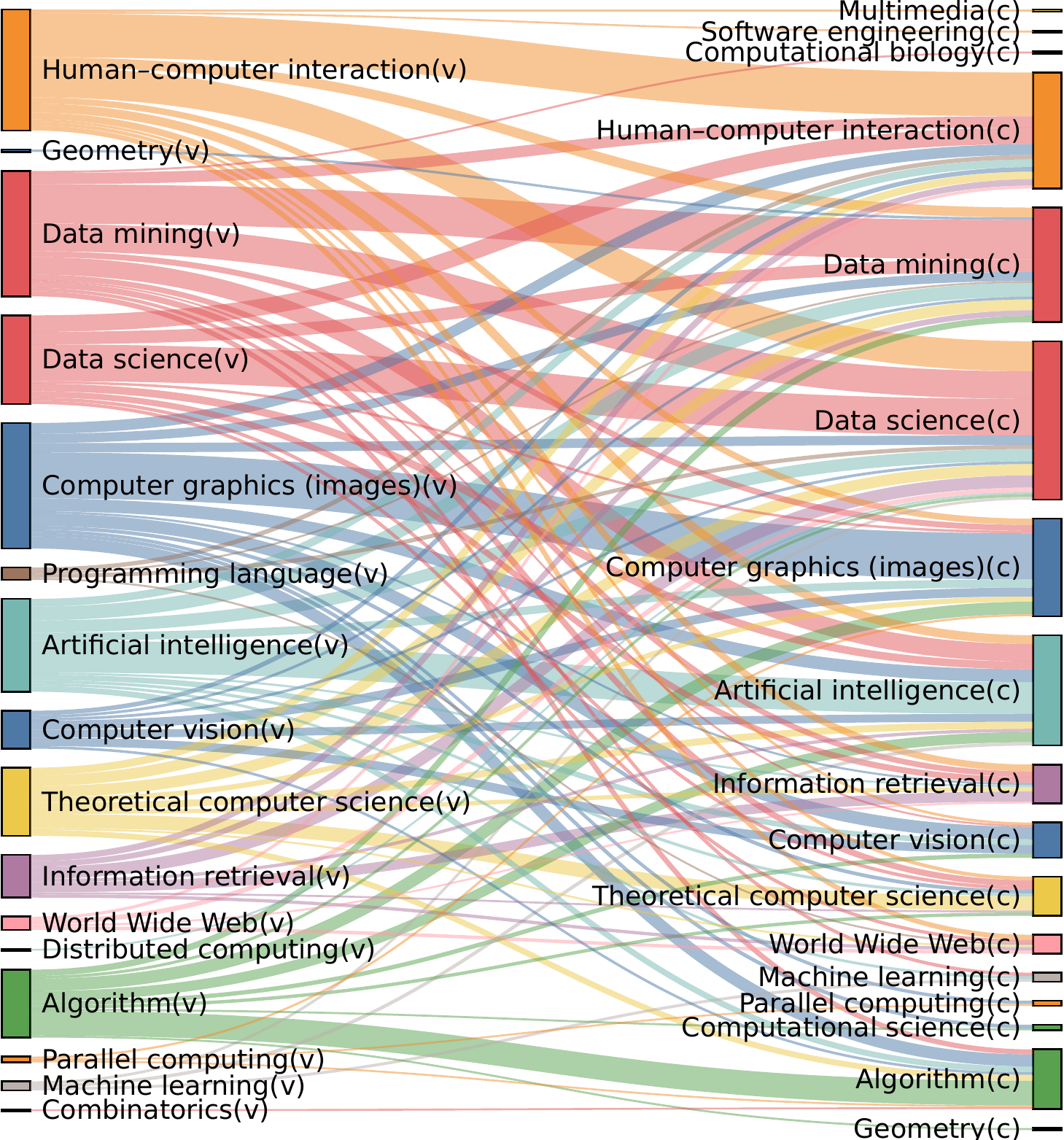}
    \caption{Citation flows based on L1 fields of study. \textbf{Left}: Top 100 citation flows, or ``pairs'', from referenced papers, i.e., papers referenced in VIS, to VIS papers; \textbf{Right}: Top 100 citation flows from VIS papers to citing papers, i.e., papers citing VIS. The letter after each concept indicates the source: r means reference, v means VIS, and c means citing papers.}
\label{fig:sankey}
\vspace{-1.5em}
\end{figure*}

Another important dimension of author affiliations is their types. There are eight affiliation types as defined by OpenAlex, which were based on the terminology set by the Research Organization Registry~\cite{ror} (ROR). Fig.~\ref{fig:author} (e) shows the proportional distribution of the affiliation types for 12,428 VIS authors. Our data revealed that 77.1\% of the authors are affiliated with a university (i.e., marked as ``education"), with the remaining 22.9\% of them scattered among companies (10.4\%), facilities (8.7\%), government (1.9\%), healthcare (1.1\%), NGOs (0.5\%), archive organizations (0.2\%) and other types (0.1\%). Because of the dominance of educational affiliations and also because of possible overlaps in the classification of non-educational affiliations, for example, some research centers could be either considered as facility, nonprofit, or even company, we decided to collapse all non-educational affiliations into one type (``non-education'') in all of our analyses that follow.

The dominance of universities did not start from the very beginning. In fact, as can be seen in Fig.~\ref{fig:author} (f), which shows the proportion of educational versus non-educational affiliations, authors affiliated with a non-educational research institution were a salient part of the conferences. For the first three years, these authors were increasing in proportion to a point where there were more authors from non-educational affiliations than those from educational ones in 1992. However, since then, authors from non-educational affiliations have steadily declined in proportion. For almost every year starting from 1995, authors affiliated with non-educational entities accounted for only 20\%-30\% of all participating authors. Since 2010, this number has fluctuated at around 20\%. In 2021, 83.6\% of all authors came from universities; only 16.4\% of them were from non-educational affiliations. 

This decline in the proportion of non-educational affiliations was due to the fact that both within the US and globally (Fig.~\ref{fig:author} (g)), the number of authors from non-educational affiliations remained stable at around 100 whereas the number of authors affiliated with universities grew steadily. In fact, as is evident in Fig.~\ref{fig:author} (g), the majority (62.8\%) of authors affiliated with non-educational entities came from the US, whereas there has been an influx of non-US authors who grew at a faster rate as a group than US authors and who mostly (80.1\%) came from universities. 

\subsection{Fields of study}

\begin{table}[!b]
\vspace{-1em}
\small
\centering
\caption{\label{tab:concepts}Availability and counts of L0-L3 concepts for VIS, referenced, and citing papers. For each source, the number in the parenthesis indicates its total count. For example, there were 3,240 VIS papers. On the right panel, the first row indicates the total number of unique concepts present at that level in that source. The second and third line indicates the percentage of papers having at least one concept in that level, and that of those having more than one. For example, 17 L0 concepts were present in VIS papers. Among 3,240 VIS publications, 99.9\% of them had at least one L0 concept, and 8.3\% of them had more than one L0 concept.}
\begin{tabular}{l|lcccc}
\toprule
Source & Description & L0 & L1 & L2 & L3 \\
\midrule
VIS papers & unique concepts & 17 & 122 & 2,036 & 730 \\(3,240) & at least one & 99.9\% & 97.3\% & 99.5\% & 83.9\% \\ & more than one & 8.3\% & 81.6\% & 97.9\% & 54.8\% \\
\midrule
Papers referenced in VIS & unique concepts & 19 & 274 & 7,905 & 5,281 \\(39,817) & at least one & 99.6\% & 95.5\% & 96.3\% & 68.0\% \\ & more than one & 19.6\% & 76.5\% & 90.2\% & 37.1\% \\
\midrule
Papers citing VIS & unique concepts & 19 & 274 & 8,140 & 5,333 \\(60,272) & at least one & 99.8\% & 96.5\% & 96.9\% & 69.1\% \\ & more than one & 15.3\% & 77.9\% & 91.9\% & 38.2\% \\
\bottomrule
\end{tabular}
\end{table}

Our fields of study analyses were based on OpenAlex's concepts data on VIS papers and their referenced and citing papers. OpenAlex assigned concepts of different levels to each paper based on the paper title, journal title, document type (e.g., journal, book, conference, patent, thesis, etc.), and the paper abstract. Within each level, the number of concepts assigned to a paper can be zero or above one. It is possible, therefore, that one VIS paper was assigned not a single Level 0 (L0) concept whereas another one had two L0 concepts. 

We first analyzed L0 concepts of VIS, referenced, and citing papers. L0 is the top level, and concepts at this level, for example, Computer Science, Mathematics, and Philosophy do not have ancestors. 

Among 3,240 VIS papers, only two missed L0 concepts, and 270 (8.3\%) of them had more than one L0 concept. There were 39,817 unique referenced papers accumulated over the past 32 years, among which 39,650 (99.6\%) of them had at least one L0 concept, and 7,795 (19.6\%) of them had more than one L0 concept. These two figures for 60,272 citing papers were 60,128 (99.8\%) and 9,201 (15.3\%), respectively. In sum, almost all papers we studied were assigned at least one L0 concept with some of them having more than one. Recognizing that some papers were interdisciplinary, we decided to include all assigned L0 concepts of each paper. Detailed statistics are reported in Table \ref{tab:concepts}.

The L0 concept of Computer Science appeared in 3,189 (98.4\%) of all 3,240 VIS publications, followed by Mathematics (185; 5.7\%), Physics (41; 1.3\%), Geology (30; 0.9\%) and Materials Science (16; 0.5\%). Similar patterns were found in referenced and citing papers. Among 39,817 unique referenced papers, the L0 concept of Computer Science appeared in 32,513 (81.7\%) of them, followed by Mathematics (5,466; 13.7\%), Psychology (2,575; 6.5\%), Physics (1,317; 3.3\%), and Medicine (1,056; 2.7\%). Among 60,272 unique citing papers, 55,008 (91.3\%) of them had the Computer Science component, followed by Mathematics (4,673; 7.8\%), Biology (1,384; 2.3\%), Medicine (1,252; 2.1\%), Geography (1,249; 2.1\%), and Psychology (1,251; 2.1\%). What these statistics reveal is that \textbf{VIS papers were mostly about, built upon, and impacting, Computer Science and Mathematical studies}. Psychology was also an important source of inspiration (For details on L0 concepts in VIS, referenced, and citing papers, refer to \textit{Supplementary Material Results 2.2}). Because the majority of L0 concepts in VIS, referenced and citing papers were all Computer Science, we focused on L1-L3 concepts in our analysis of (1) the popularity trends of concepts in VIS publications and (2) the citation flows based on concepts. 

As clear in Fig.~\ref{fig:teaser}, at each level, VIS publications over the years were concentrated in only a few concepts. 122 L1 concepts were present in VIS papers but only 6 of them appeared more than 100 times throughout the 32 years. There were significantly more L2 (2,036) and L3 (730) concepts present in VIS publications, but again, only 27 (for L2) and 8 (for L3) of them had a frequency of over 100, respectively. Not surprisingly, except for three concepts in L2 (Context, Domain, and Focus) and one in L3 (Representation)~\footnote{Flow visualization is not considered a subfield of Computer Science by OpenAlex, which we considered to be a mistake.}, all of these high-frequency concepts were subfields within the discipline of Computer Science. We also noticed that the overall trends in science were reflected in the ups and downs of popular concepts in VIS publications such that we saw an increase in the number of publications involving Artificial Intelligence (AI), Data Mining, Human Computer Interaction (HCI), Data Science, Machine Learning (ML) and Analytics whereas the popularity of traditional subfields, i.e., Computer Graphics and rendering had been declining.

We were also interested in which (sub)fields VIS was built upon and where VIS influences flew. Within each level, we chose the concept with the highest score to represent each paper if there were multiple concepts assigned; Otherwise, citation flows would be over-represented by interdisciplinary papers. \textbf{We found that at each level, citations mostly flew between the same set of subfields.} For example, in L1, VIS papers mostly cited, and were mostly cited by, studies falling into AI, HCI, Computer Graphics, Algorithm, Data Science, Data Mining, Theoretical Computer Science, and Information Retrieval, as can be seen in Fig.~\ref{fig:sankey}. Similar patterns were found in L2 and L3 concepts. More details on, and temporal patterns in, citation flows are available in our interactive visualizations at \href{https://32vis.hongtaoh.com/}{\texttt{https://32vis.hongtaoh.com}}. 

Visualization as a field is interdisciplinary \cite{isenberg2016visualization, sarvghad2022scientometric}. We explored the co-occurrence of concepts within each level for VIS publications. We found that at each level, co-occurrences of concepts were mostly between the most frequently appearing concepts. This finding, that \textbf{VIS is interdisciplinary, but only confined to a few fields, both within (L1-L3) and outside of (L0) Computer Science}, corroborates and adds to the result in \cite{sarvghad2022scientometric}. More details on concept co-occurrences of VIS papers can be found in \textit{Supplementary Material Results 2.3}.

\begin{figure}[!b]
\centering
\vspace{-1.0em}
\includegraphics[scale=0.27]{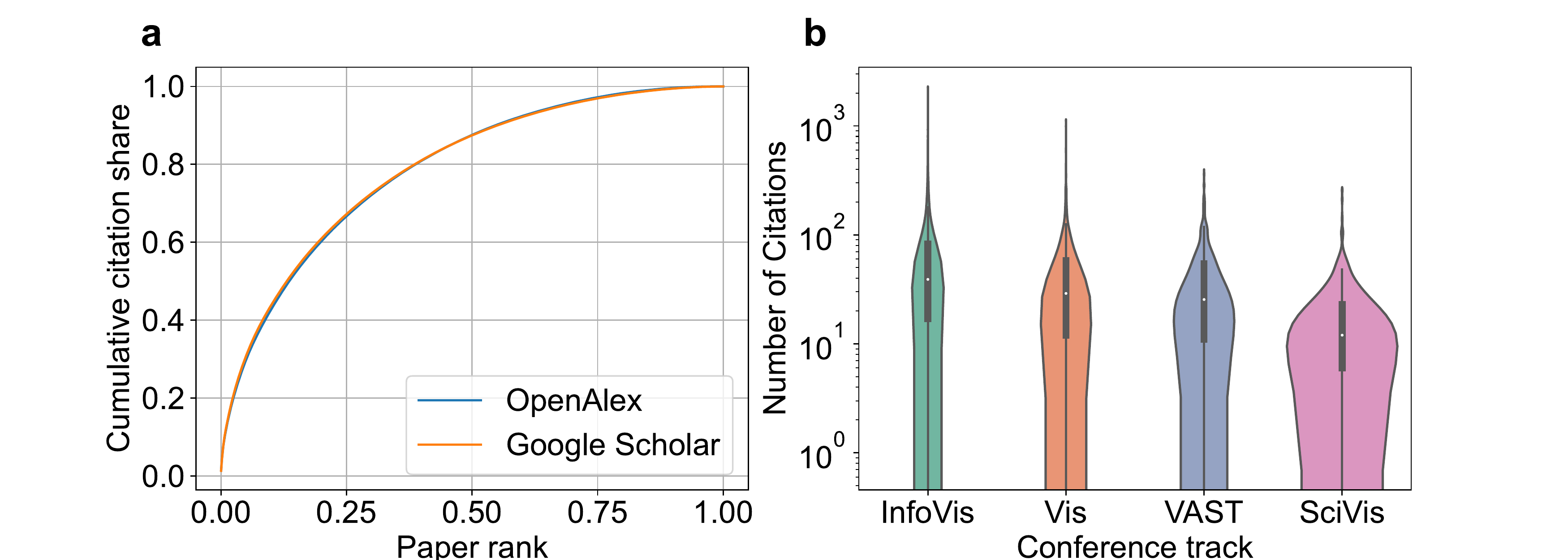}
\includegraphics[scale=0.3]{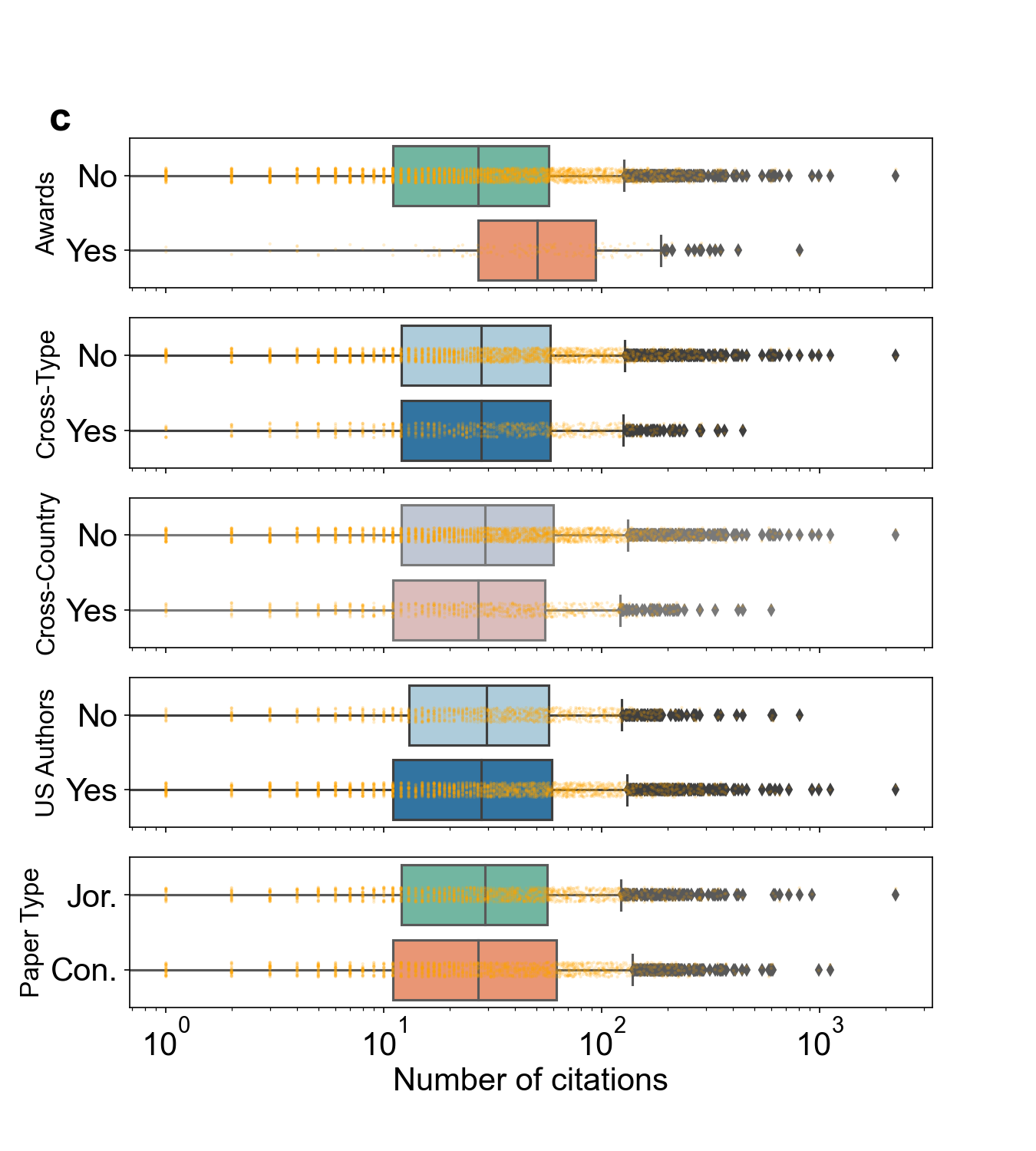}
\vspace{-1em}
\caption{\label{fig:citations} \textbf{a} shows citations were highly skewed such that the top 20\% of papers received 60\% citations. \textbf{b} are violin plots displaying the distribution of citations within each conference track. \textbf{c} compares citations of VIS papers under different categories. Note that the time factor may distort interpretations of \textbf{b} and \textbf{c}. For example, SciVis is the youngest track and this might impact its citations. This is why we built a multiple regression model in \ref{cit_analysis}: to consider all known (and available) factors that might influence citation counts.}
\end{figure}

\subsection{Citation analysis}
\label{cit_analysis}

Like in other bibliometric studies~\cite{seglen1992skewness, albarran2011skewness, bornmann2017skewness}, citations of IEEE VIS papers were heavily skewed to the right such that the top 20\% papers received 60\% of all citations, as is shown in Fig.~\ref{fig:citations} (a). This pattern was the same if we used citation data from Google Scholar, which is closely correlated with that from OpenAlex (with a Spearman's rank correlation of 0.95). 

We were interested in which paper characteristics were associated with more citations. We first compared citation counts group-wise, for example, between cross-type and non cross-type collaboration papers, using non-parametric tests (We used one-way ANOVA for four different conference tracks). Our results (Fig.~\ref{fig:citations} (b) and (c)) showed that only award-winning and conference tracks were significant predictors. \textit{Supplementary Material Results 2.4} covers more details.

We then regressed the number of citations on year distance from 2020, conference track, paper type, number of authors, cross-type collaboration, cross-country collaboration, involving US authors or not, and award-winning. Following the practice of \cite{bartneck2009scientometric}, we excluded VIS2021 papers in our analysis because it was too early for them to receive citations. We made sure the multicollinearity of the model was not an issue. Linearity and homoscedasticity had some issues, which could be solved by performing log10 transformation on citation counts. Doing so, however, would make the interpretations of our results less intuitive, so we decided to still use the raw citation counts.  

Regression analyses showed that publication year, conference tracks, paper types and awards are significant predictors of citation counts such that recent publications ($b = -2.51$, $t = -6.78$, $p < .001$), and papers presented in SciVis ($b = -42.75$, $t = -7.57$, $p < .001$), Vis ($b = -37.25$, $t = -8.67$, $p < .001$) and VAST ($b = -14.60$, $t = -2.97$, $p < .01$) had significantly lower citations. Having won an award ($b = 28.18$, $t = 4.20$, $p < .001$) or been published in a journal rather than a conference proceeding ($b = 22.25$, $t = 4.57$, $p < .001$) brought significantly more citations. Number of authors, cross-type collaboration, cross-country collaboration, and involving US authors did not significantly correlated with the number of citations a paper received. Note that although this model is significant ($F(10,3059) = 18.45, p < .001$), it only explains $5.4\%$ (adjusted $R^2 = .0537$) of the variance in citation counts. Regressing number of citations from Google Scholar, or performing log10 transformation on OpenAlex citations did not change the overall results. (For detailed analyses and regression results, see \textit{Supplementary Material Results 2.5}).

\section{Discussion}

We began our study with two questions: (1) where VIS authors are from, and (2) where VIS stands in science. To answer these questions, we collected VIS paper metadata, and data on VIS authors and citations from vispubdata.org, IEEEXplore, and OpenAlex, among other sources.

We found that the popularity and impact of VIS have been increasing. The number of accepted publications and that of unique participating authors have been increasing. Given the relatively stable acceptance rate of 25\% at VIS during the past decade, these increases indicate there has been a growing number of people involved in visualization research. We also found that a growing number of non-VIS studies have been citing VIS papers. We did not know the overall growth rate for all non-VIS papers (whether citing VIS or not), but this trend nonetheless implied that research outside of the VIS community was influenced by VIS.

Collaborations within VIS have been becoming increasingly popular. Sarvghad et al.~\cite{sarvghad2022scientometric} explored inter-institutional and interdisciplinary collaborations in VIS and found that both numbers have been growing in the past 30 years. We corroborated this finding from different aspects. Our data showed that collaborations between authors from different countries, and between authors from different affiliation types, i.e., education and non-education, experienced rapid growth. Cross-country collaboration grew from 4\% in 1990 to 45\% in 2021, and cross-type collaboration increased from 10\% in 1990 to 32\% in 2021. Our regression analyses showed, however, neither cross-country collaboration nor cross-type collaboration yielded more citations.

Even though cross-country collaboration has been increasingly popular, we found that these collaborations were concentrated in a few countries. The top ten most collaborative countries (US, Germany, China, Austria, UK, France, Canada, Netherlands, Switzerland, and Australia) were present in 98\% of all collaboration pairs, and collaborations among these most active countries were responsible for 71\% of all cross-country collaborations throughout the history of VIS. This is not surprising; after all, these countries were also the top ten most important author sources from where 93\% of all VIS authors came.

VIS authors were also concentrated in a few countries and continents, with some changes in author shares over the years. Authors from the US dominated VIS in the first decade, then authors from Germany became a significant part in the second decade of VIS. In the third decade, the number of authors from China increased drastically. In terms of total author counts, the US (52.9\%) is still the largest author source, followed by Germany (13.3\%) and China (8.2\%). If we consider year by year statistics, China replaced Germany as the second largest author source in 2017 and has been in that position ever since. Although US authors have been declining steadily in terms of proportion, they still had disproportionate influences on VIS. In the past decade, we saw an upward trend in the percentage of papers involving at least one author from the US. In 2021, although US authors accounted for 42.6\% of all authors, they were present in 58.8\% of all VIS2021 publications.

The results regarding the proportion and participation of non-educational affiliations are interesting. We found that the proportion of collaboration between authors from universities and those from non-educational affiliations (e.g., companies, governments, NGOs, etc.) showed an upward trend: in 1990 only 10\% of all publications were cross-type collaborations whereas this number grew to 32\% in 2021. Given this trend, it is surprising that the proportion of authors from non-educational affiliations, both globally and within the US, has been declining steadily. In the year of 1992, 59\% of VIS authors were affiliated with non-educational entities whereas this number plunged to 16\% in 2021. We concluded that this was because of the stable number of authors from non-educational entities at around 100 each year amidst a steadily growing number of authors from universities both globally and within the US. The fact that cross-type collaborations were rising in popularity whereas the proportion of authors from non-educational affiliations was declining indicates that the small number of these authors were actively participating in VIS projects every year.

Using different data and from different perspectives, we corroborated the findings in \cite{sarvghad2022scientometric} that (1) most VIS papers fell into the discipline of Computer Science, and that (2) interdisciplinary collaboration within VIS was confined to only a few disciplines. Our data showed that the L0 concept of Computer Science was present in all except two of 3,240 VIS papers. Almost all of the high-frequency concepts in L1, L2, and L3 were subfields of Computer Science. We found that co-occurrences of concepts within each level were only between a few (sub)fields. Regarding the role of VIS in science, we found that the scope of topics, inspirations, and impacts of VIS research are limited. First, VIS papers, and their referenced and citing papers mostly fell into the domain of Computer Science and Mathematics. Time-series analyses (see \href{https://32vis.hongtaoh.com/}{\texttt{https://32vis.hongtaoh.com}}) show that this pattern is stable throughout the past 32 years. Second, within each level of fields of study, citations mostly flew between the same set of (sub)fields, indicating that VIS publications did not have diversified inspirations and impacts.

Our regression analyses showed that citations counts were significantly lower for papers published in recent years, and papers presented at SciVis, Vis and VAST. Journal papers and award-winning papers had significantly more citations. This contradicted the finding in \cite{bartneck2009scientometric} that award-winning CHI papers did not receive more citations than a randomly selected publication. Our results implied that awards at VIS were able to identify high-impact works.

\section{Limitation and future work}
Our study is not devoid of limitations. The quality of OpenAlex Concepts data may not be desirable for more granular concepts, i.e., L2 and L3. For example, as Fig.~\ref{fig:teaser} shows, there is a L3 concept of ``Representation (politics)'', which is unlikely for VIS publications. That said, considering that (1) most concepts shown in Fig.~\ref{fig:teaser} appear related to what VIS publications are about; and (2) the general trends mirror what is happening in the real world such that AI, ML, HCI and Data Science are becoming popular, we believe the quality of OpenAlex concept tagging is reliable, especially for L0 and L1, from which most of our conclusions on fields of study were drawn. Also, the hierarchy of OpenAlex concepts may not be optimal. This is because OpenAlex's concept tagging is trained on MAG data that only manually inspected the hierarchical relationship between L0 and L1 concepts~\cite{Shen2018AWS}. This drawback, however, has minimal impact on our conclusions because (1) our conclusions were mostly based upon L0 and L1 concepts, and (2) for L2 and L3 concepts, which are concentrated in only a handful of high-frequency concepts for VIS papers, we have manually checked whether they belong to Computer Science.

Our study has the potential to inspire future work in several directions. First, OpenAlex only launched itself in early 2022 and has not been widely used in scientific studies. We found that their data were reliable and can be integrated with other data sources. Other researchers may be motivated by our work and consider OpenAlex data, and our workflow, in their studies. Second, analysis of citations flows based on fields of study is a useful tool to identify the academic bases (i.e., which fields are they built upon) and impacts (i.e., which fields are they influencing) of a scholarly field. Our approaches can be applied to analyses of many other fields, such as HCI and CSCW. Last but not least, our present study did not include conference papers in PacificVis and EuroVis and journal papers in \textit{Journal of Visualization} and \textit{Information Visualization}. Future work may include those publications to draw a more complete picture of the field of Visualization.

\acknowledgments{
We thank Jason Priem and Richard Orr from OpenAlex for their help in our data collection. We would like to extend our sincere gratitude to Ruobin Wang for her help in creating the diagram for the data processing pipeline where Yaxin Hu also contributed. We thank Attila Varga, Jisung Yong, and Munjung Kim for their feedback on our study designs, and Sadamori Kojaku, Chaoqun Ni, and Lili Miao for their initial suggestions on data sources. We thank Yong-Yeol Ahn, Filipi Silva, and Larry Zhang for their help. We also thank Christian Schulte zu Berge, Bernice Rogowitz, Isaac Cho, David Laidlaw and Vadim Slavin for responding to our emails regarding author affiliations.}

\bibliographystyle{abbrv}
\bibliography{main}
\end{document}



\title{Supplementary material for \vspace{1em}\\
  \large Thirty-Two Years of IEEE VIS: \\
    Authors, Fields of Study and Citations} 
\date{\today}
\author{
  Hongtao Hao\\
  \texttt{hongtaoh@cs.wisc.edu}
  \and
  Yumian Cui\\
  \texttt{ycui53@wisc.edu}
  \and
  Zhengxiang Wang\\
  \texttt{jackwang196531@gmail.com}
  \and
  Yea-Seul Kim\\
  \texttt{yeaseul.kim@cs.wisc.edu.}
}

\maketitle 

\tableofcontents


\section{Methods}

\subsection{Get DOIs for VIS 2021 papers}

For a paper whose DOI as obtained from Crossref contains the IEEE prefix, i.e., \texttt{10.1109}, and whose title as obtained from Crossref matches the queried title, we regarded the query result as correct. There are twenty-seven papers whose DOI contained this prefix, but whose title as obtained from Crossref ID did not match that on VIS 2021. By comparing these titles manually, we found that four of them were indeed mismatches; the rest were correct results with insignificant variations in titles. There were twenty-three papers whose DOIs as obtained from Crossref did not contain \texttt{10.1109}; Their query results were obviously wrong. In sum, 23 + 4 = 27 papers had incorrect query results. We manually collected their DOIs from IEEE Xplore.

\subsection{Choose the right data sources}

We first ruled out JSTOR and PubMed. JSTOR does not contain reference and citation information at all. PubMed has impressive coverage on VIS papers; IEEE Xplore even links each publication to its PubMed page. Unfortunately, PubMed provides very limited data on authors and citations and provides no data on references at all.

We then examined the Web of Science and Scopus. IEEE Xplore displays each publication's citation metrics as measured by these two databases, whose coverage on VIS papers, however, is not desirable. We randomly selected 100 papers from the 3,242 VIS papers and collected their citation metrics data as displayed on IEEE Xplore. Only 71 of them had citation counts from Scopus; this figure for Web of Science was 68. We validated this result by identifying VIS papers via DOI query on Web of Science. Among all 3,242 papers, 25\% were not identifiable. Some of the inaccessible papers could be identified by title query but this approach is not only labor-intensive and error-prone but also hard to automate and reproduce. Apart from this low coverage on VIS papers, we gave up Web of Science and Scopus also because of their paywalls. Both are proprietary and therefore not easily accessible. Even if we collect information from them through web-scraping, we are not allowed to share our data publicly, which defies our wish to make our work reproducible.

Crossref is free and has surprisingly good coverage on VIS papers. Among all 3,242 papers, only one is missing from its database. A closer examination of their data, however, revealed Crossref is not an ideal source of data for our present study either. For all the first authors, 99\% of them miss affiliations, rendering Crossref useless for author-related data collection. In addition, Crossref provides information on fields of study only at the journal level rather than at the paper level. This means that on Crossref, VIS papers either miss fields of study, or ``subject'' as Crossref codes it, or they have exactly the same fields of study, which are assigned to the journal of \textit{IEEE Transactions on Visualization and Computer Graphics}.

Semantic Scholar \footnote{https://www.semanticscholar.org/} provides a powerful API. It provides citation and reference counts and titles of both referenced and citing papers. What it lacks, however, is the author data and the fields of study. Because these two types of data are important in our study, Semantic Scholar is not a viable choice.

This left us only three options: Google Scholar, Microsoft Academic Graph (MAG), and OpenAlex \footnote{https://openalex.org/}. Google Scholar is the most popular database in terms of web traffic \footnote{https://blogs.lse.ac.uk/impactofsocialsciences/2021/05/27/goodbye-microsoft-academic-hello-open-research-infrastructure/}. It provides the citation counts, which we collected, but no other useful information for this study. Between MAG and OpenAlex we chose the latter mostly because Microsoft discontinued MAG recently \footnote{https://www.microsoft.com/en-us/research/project/academic/articles/microsoft-academic-to-expand-horizons-with-community-driven-approach/}. If we get our data from MAG, it is very likely that our workflow will not be useful for future researchers. Like MAG, OpenAlex is a scholarly database freely accessible to the public. A large amount of its data is directly from MAG. OpenAlex is new but growing and well maintained. One special advantage of OpenAlex over MAG for our study is that it provides author affiliations' type, i.e., education, company, government, etc., which is not available on MAG.

\subsection{How OpenAlex collects concepts data}

OpenAlex uses the word ``Concept'' to describe fields of study. OpenAlex built a multi-class classifier based on MAG’s fields of study data to assign concepts to each publication indexed in OpenAlex. OpenAlex employs 65K concepts taken from \url{wikidata.org}. These concepts have six levels. There are 19 Level-0 concepts, each indicating the highest level concept, for example, computer science, mathematics, medicine, physics, chemistry, etc. The larger the number of the level, the more detailed this concept is. OpenAlex detailed how they operate concepts tagging in their whitepaper~\footnote{\url{https://docs.google.com/document/d/1OgXSLriHO3Ekz0OYoaoP_h0sPcuvV4EqX7VgLLblKe4/}}. They also released their codes of concept tagging~\footnote{https://github.com/ourresearch/openalex-concept-tagging}.

\section{Results}

\subsection{Changes in average number of authors per paper}

\begin{figure}[!t]
\centering
  \includegraphics[scale = 0.7]{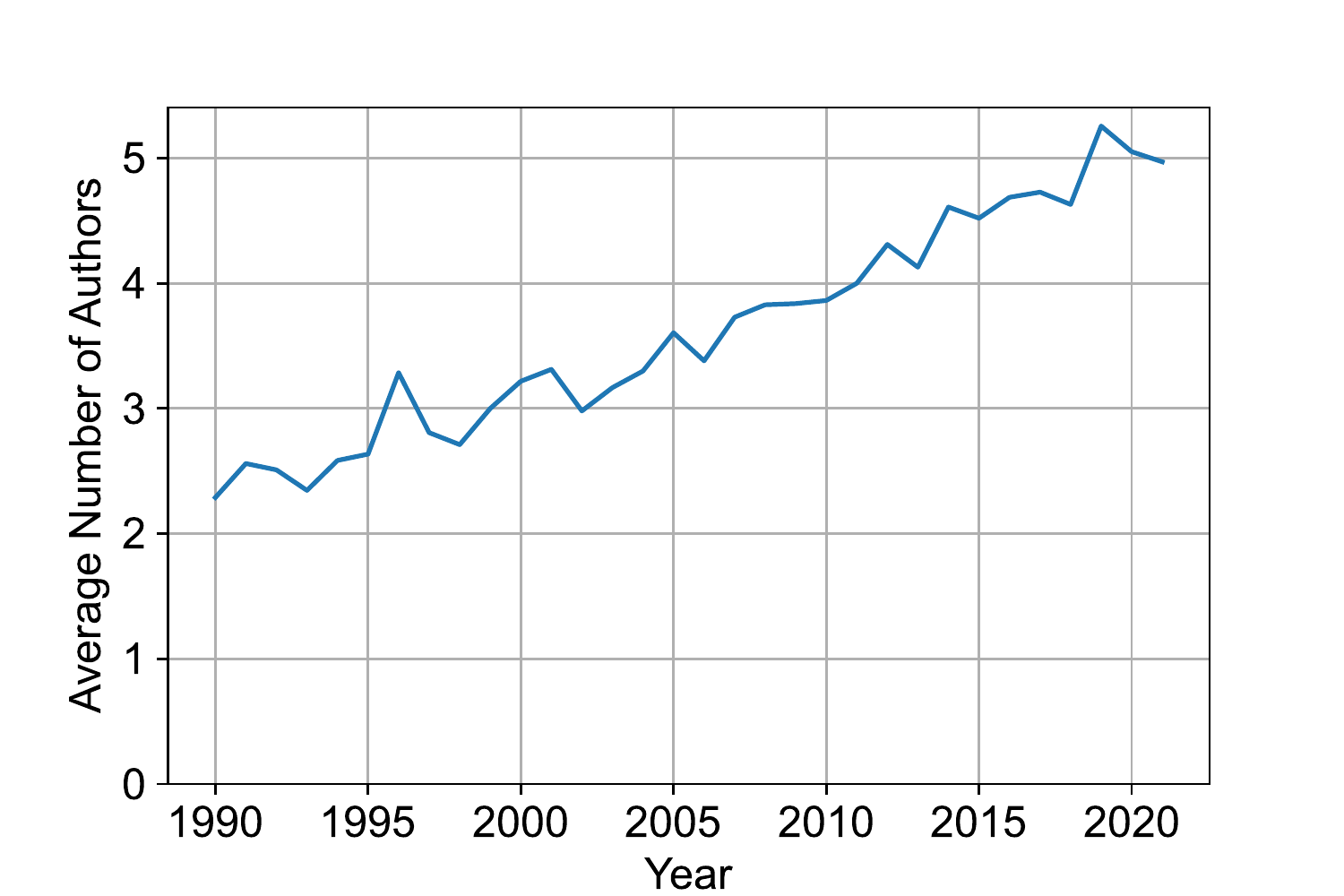}
    \caption{Average number of authors by year}
\label{fig:ave_author_num}
\end{figure}

Fig.~\ref{fig:ave_author_num} shows the changes in the average number of authors per paper by year. It is clear that it has an upward trend.

\subsection{Distribution of L0 concepts in VIS, referenced, and citing papers}

\begin{figure}
\centering
  \includegraphics[scale = 0.5]{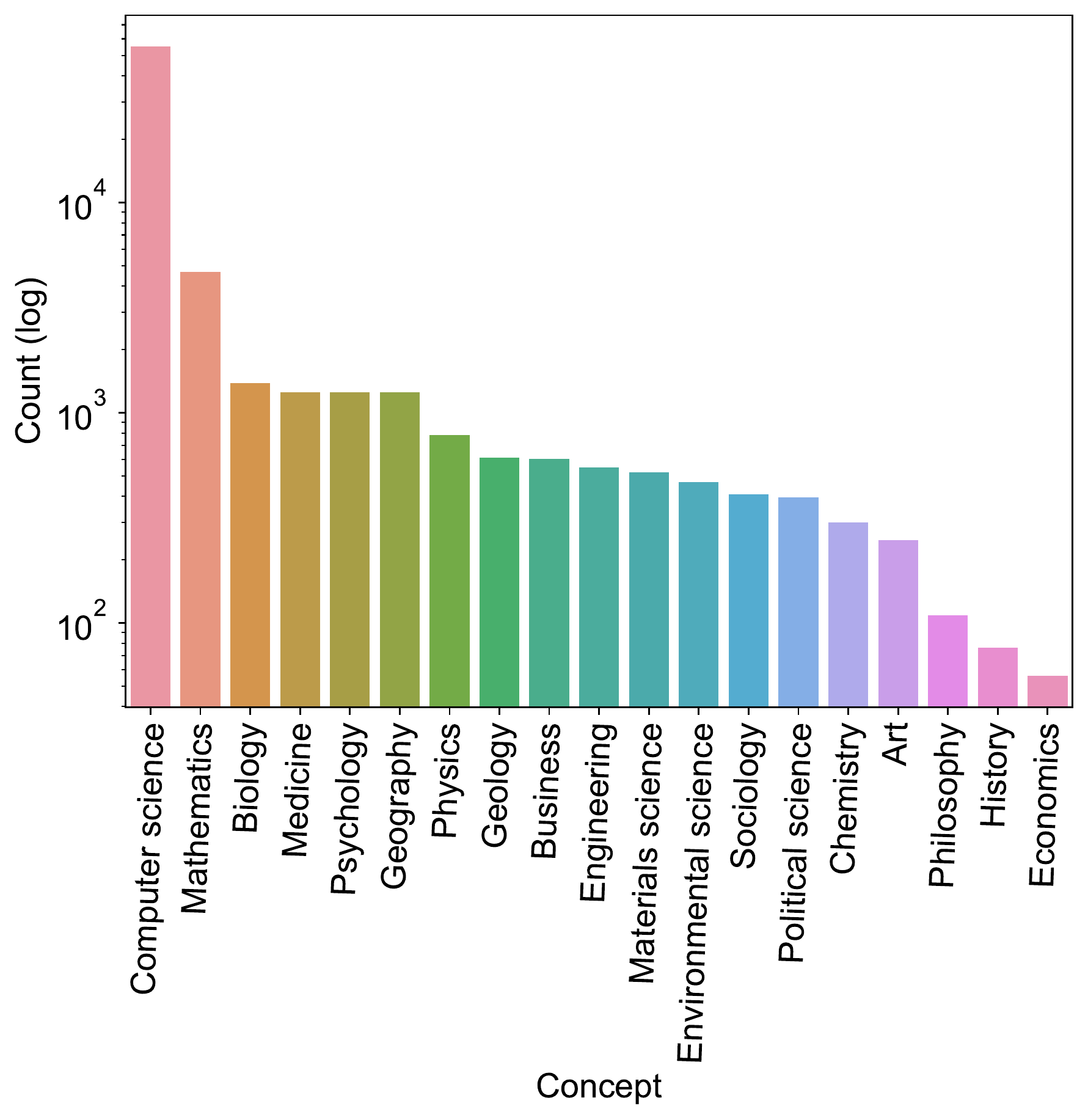}
    \caption{Distribution of L0 concepts among citing papers}
\label{fig:cit_l0}
\end{figure}

The following figures, namely Fig. \ref{fig:vis_l0}, Fig. \ref{fig:ref_l0}, and Fig. \ref{fig:cit_l0} present the distribution of Level 0 Concepts among VIS, referenced, and citing papers. From these figures, it is clear that in all of them, Computer Science is the most salient part, followed by Mathematics.

\begin{figure}[!t]
\centering
  \includegraphics[scale = 0.5]{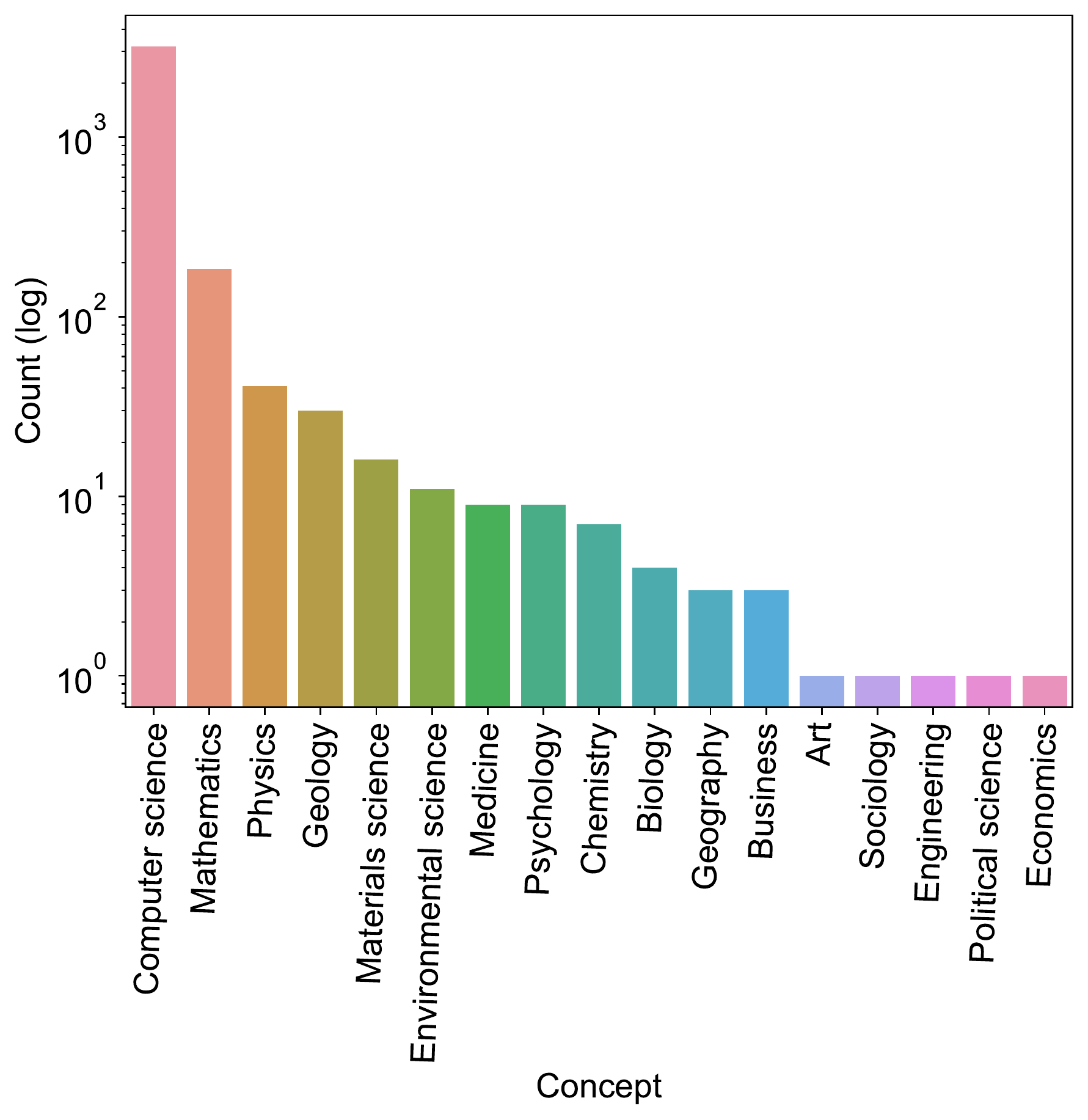}
    \caption{Distribution of L0 concepts among VIS papers}
\label{fig:vis_l0}
\end{figure}

\begin{figure}
\centering
  \includegraphics[scale = 0.5]{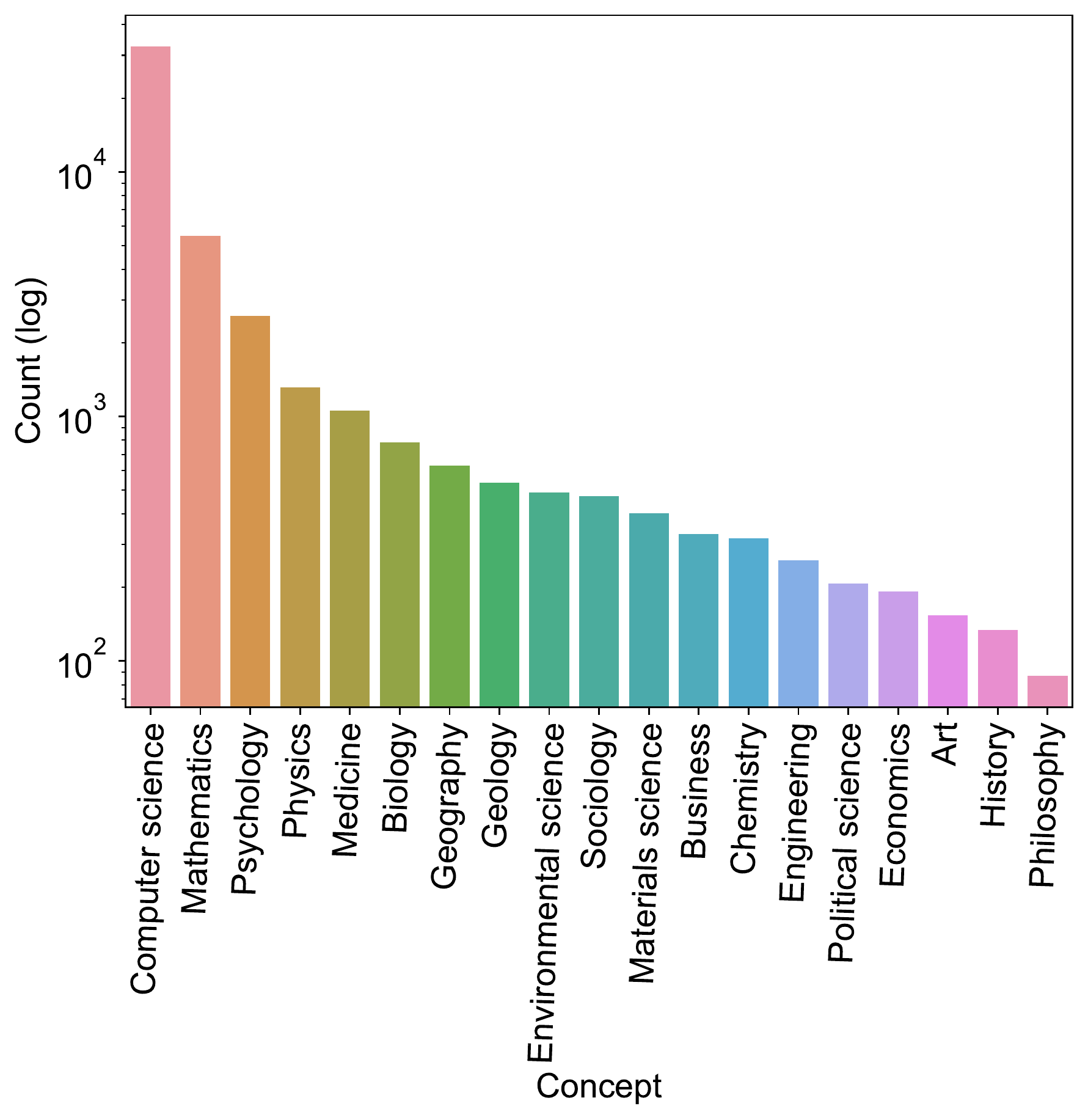}
    \caption{Distribution of L0 concepts among referenced papers}
\label{fig:ref_l0}
\end{figure}

\subsection{Concepts co-occurrence in VIS papers}

\begin{figure}
\centering
  \includegraphics[scale = 0.4]{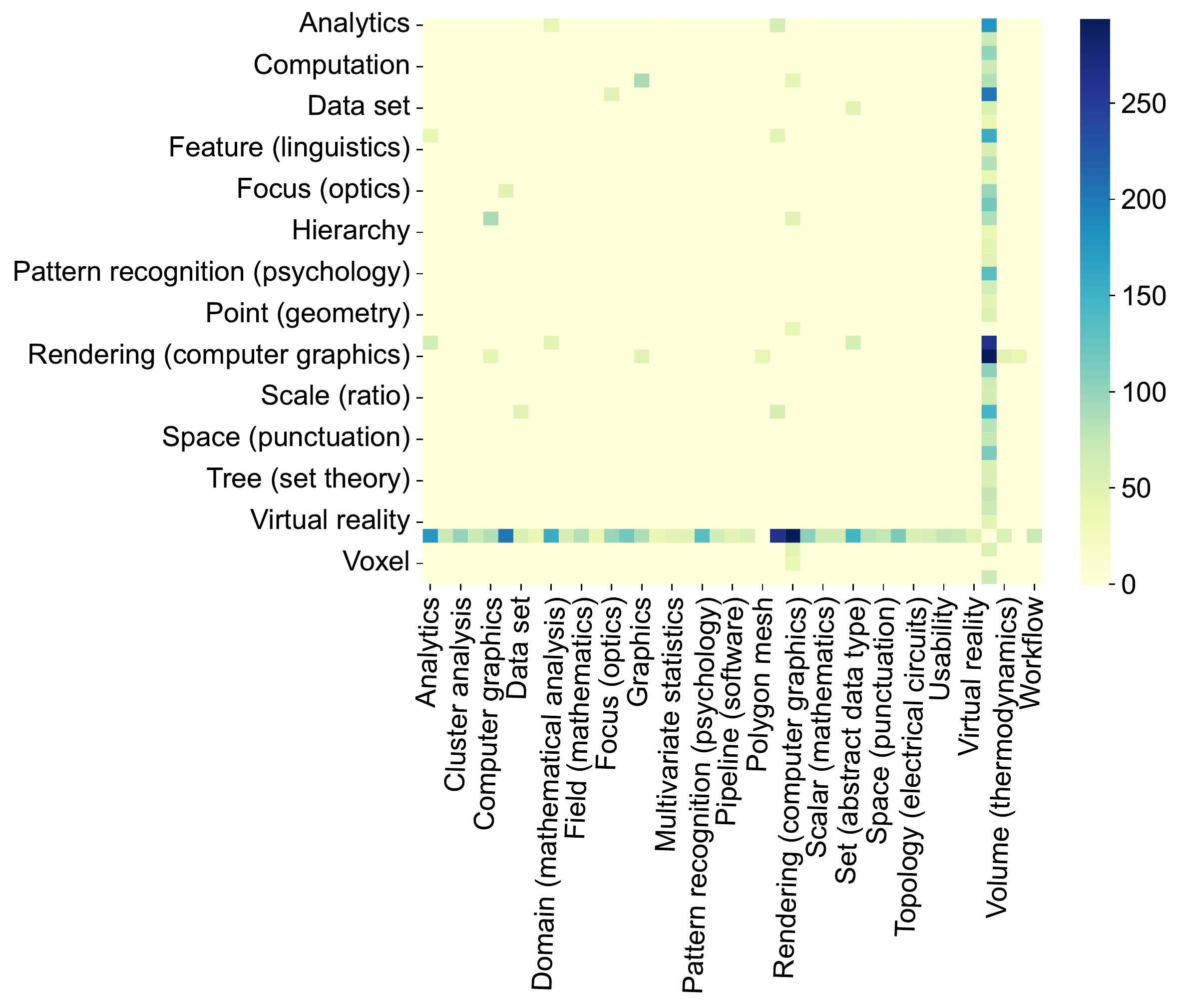}
    \caption{Co-occurrence of L2 concepts in VIS; pairs appeared at least 40 times.}
\label{fig:co_l2}
\end{figure}

\begin{figure}
\centering
  \includegraphics[scale = 0.4]{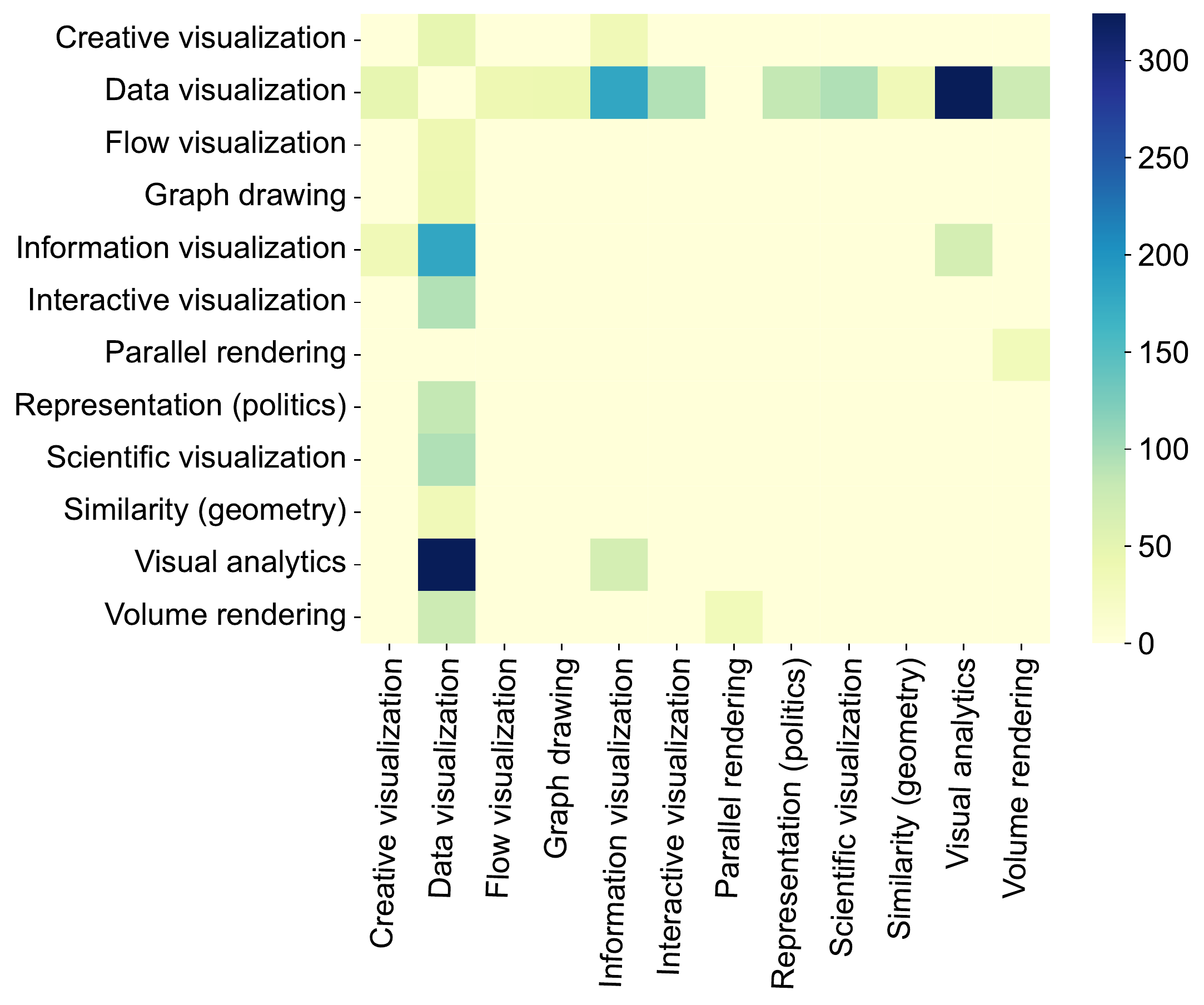}
    \caption{Co-occurrence of L3 concepts in VIS; pairs appeared at least 30 times.}
\label{fig:co_l3}
\end{figure}

The following figures, namely, Fig. \ref{fig:co_l0}, Fig. \ref{fig:co_l1}, Fig. \ref{fig:co_l2}, Fig. \ref{fig:co_l3} show which concepts (L0-L3) in VIS papers co-occurred. From these results, it is clear that at each level only a few Concepts co-occurred, with the highest-frequency Concept at each level attracting the majority of co-occurrences.

\begin{figure}
\centering
  \includegraphics[scale = 0.4]{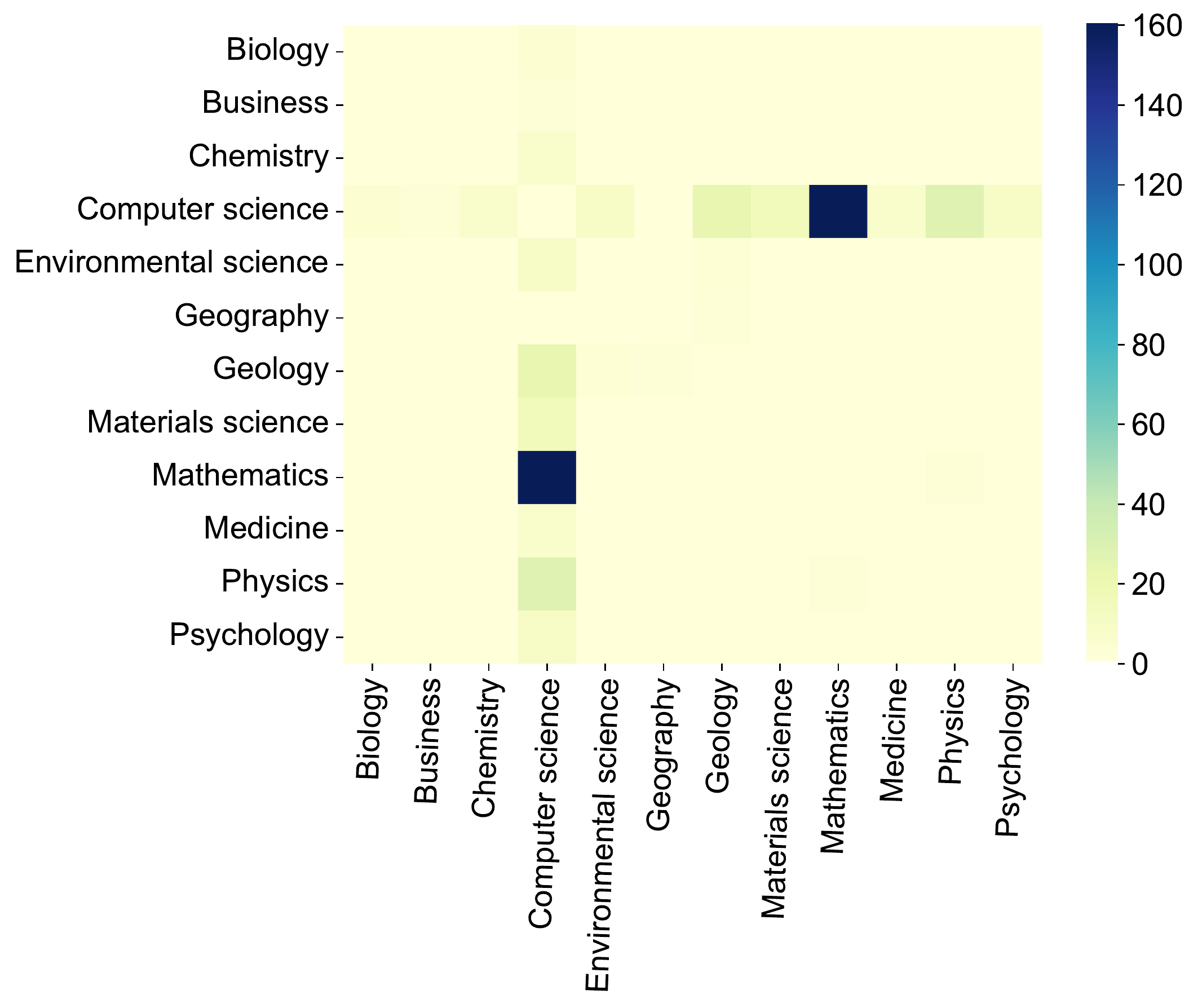}
    \caption{Co-occurrence of L0 concepts in VIS; pairs appeared at least once.}
\label{fig:co_l0}
\end{figure}

\begin{figure}
\centering
  \includegraphics[scale = 0.4]{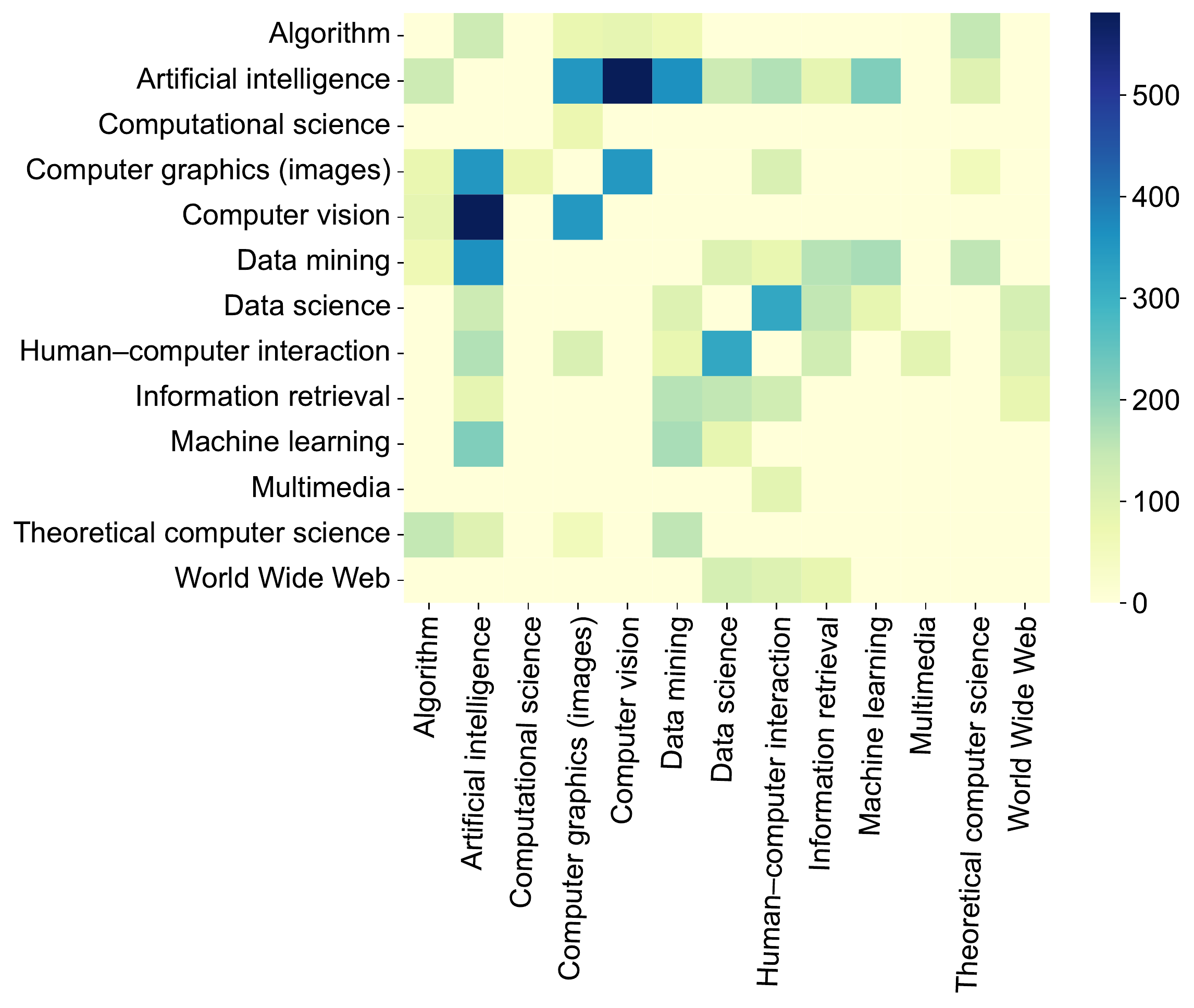}
    \caption{Co-occurrence of L1 concepts in VIS; pairs appeared at least 50 times.}
\label{fig:co_l1}
\end{figure}

\subsection{Group-wise citation analysis}

Given the highly skewed nature of citation counts, we employed non-parametric tests to investigate differences in groups. We rely on citation counts from OpenAlex, and will report results if Google Scholar citation counts yielded a different result.

Mann-Whitney tests showed that neither papers written by authors from affiliations of different types ($U = 606,271, p = 0.86$), nor those by authors from different countries ($U = 822,967, p= 0.29$), obtained more citations than their counterparts. Papers involving authors from the United States ($U = 1,058,864, p = 0.69$) did not get more impacts than their counterparts either. There were also no differences in the citations between journal and conference papers ($U = 1,201,856, p = 0.33$). Google Scholar citations, however, showed that cross-country collaboration papers got significantly fewer citations ($U = 783,863, p < 0.01$), and that conference papers got more citations than journal papers ($U = 1,078,280, p < 0.001$). Citations from both OpenAlex ($U = 292,878$) and Google Scholar ($U = 287,031$) both revealed that award-winning papers are more impactful ($p < 0.001$).

We ran a one-way ANOVA to compare the effect of conference tracks on the number of citations. Test results showed there were significant differences among groups ($F(3, 3066) = 35.19, p < 0.001$). InfoVis papers got the highest number of citations, followed by Vis, and VAST. SciVis papers were the lowest. Tukey's HSD test revealed that all group pairs had significantly different citations except for that between Vis and VAST ($p = 0.21$).  



\subsection{Regression results for citation analysis}

Regression results are as follows:

\subsubsection{OpenAlex citations}
\begin{table}[!htbp] \centering
  \caption{Regression results with OpenAlex citations. Estimates are unstandardized coefficients with standard error and $p$ values.}
  \label{}
\begin{tabular}{@{\extracolsep{5pt}}lc}
\\[-1.8ex]\hline
\hline \\[-1.8ex]
 & \multicolumn{1}{c}{\textit{Dependent variable:}} \\
\cline{2-2}
\\[-1.8ex] & Number.of.Citations \\
\hline \\[-1.8ex]
 Year.Distance.from.2020 & 2.507$^{***}$ \\
  & (0.370) \\
  & \\
 ConferenceSciVis & $-$42.747$^{***}$ \\
  & (5.646) \\
  & \\
 ConferenceVAST & $-$14.599$^{***}$ \\
  & (4.919) \\
  & \\
 ConferenceVis & $-$37.252$^{***}$ \\
  & (4.298) \\
  & \\
 PaperTypeJ & 22.250$^{***}$ \\
  & (4.866) \\
  & \\
 Number.of.Authors & $-$0.289 \\
  & (0.884) \\
  & \\
 Cross.type.CollaborationTrue & $-$1.323 \\
  & (3.630) \\
  & \\
 Cross.country.CollaborationTrue & $-$3.952 \\
  & (3.826) \\
  & \\
 With.US.AuthorsTrue & 4.049 \\
  & (3.076) \\
  & \\
 AwardTrue & 28.179$^{***}$ \\
  & (6.705) \\
  & \\
 Constant & 28.486$^{***}$ \\
  & (8.196) \\
  & \\
\hline \\[-1.8ex]
Observations & 3,070 \\
R$^{2}$ & 0.057 \\
Adjusted R$^{2}$ & 0.054 \\
Residual Std. Error & 80.168 (df = 3059) \\
F Statistic & 18.445$^{***}$ (df = 10; 3059) \\
\hline
\hline \\[-1.8ex]
\textit{Note:}  & \multicolumn{1}{r}{$^{*}$p$<$0.1; $^{**}$p$<$0.05; $^{***}$p$<$0.01} \\
\end{tabular}
\end{table}

\begin{figure}
\centering
  \includegraphics[scale = 0.8]{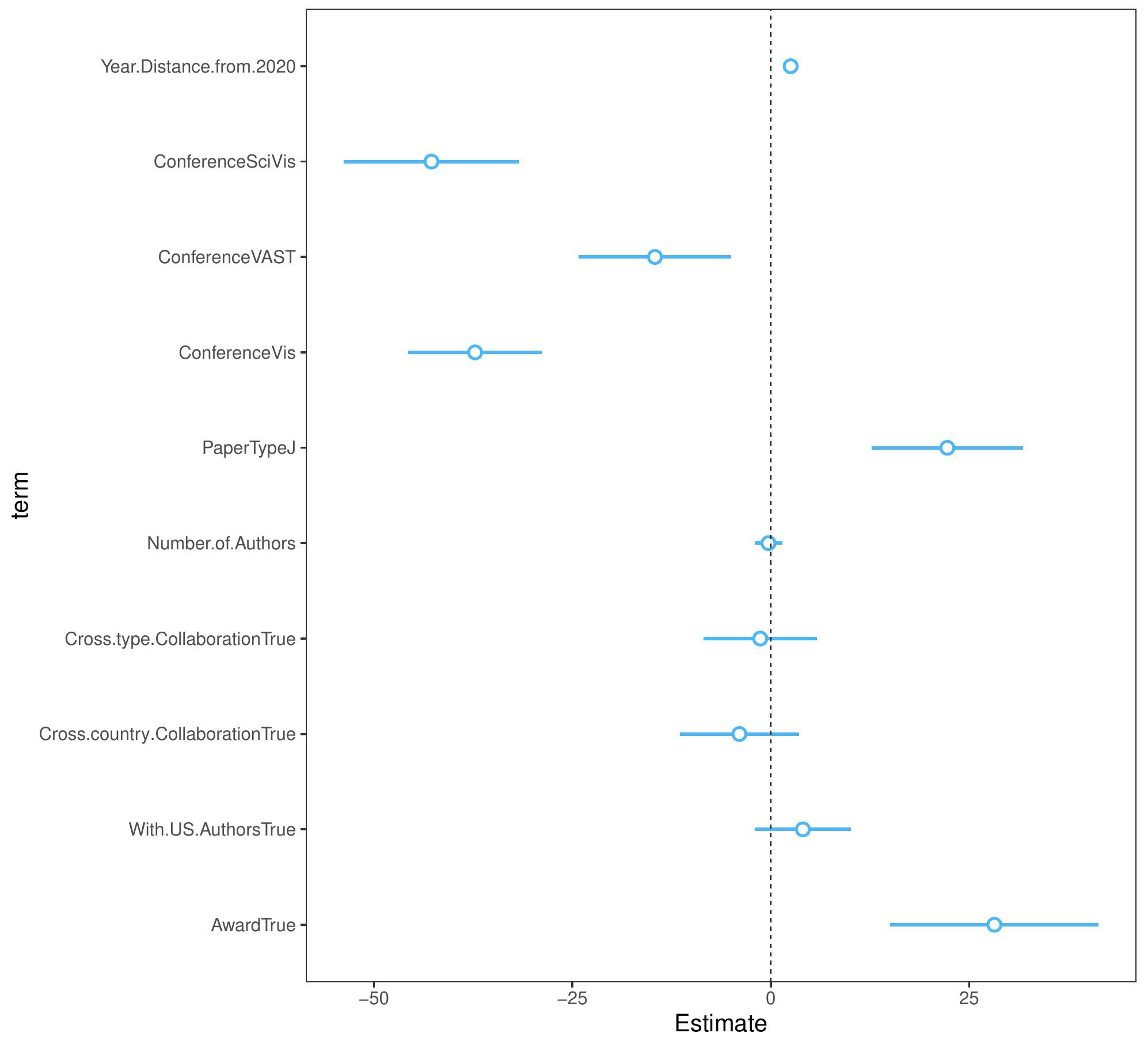}
    \caption{Regression results with OpenAlex citations. Estimates are unstandardized coefficients with 95\% confidence intervals.}
\label{fig:reg_openalex}
\end{figure}

\subsubsection{Google Scholar citations}
\begin{table}[!htbp] \centering
  \caption{Regression results with Google Scholar citations. Estimates are unstandardized coefficients with standard error and $p$ values.}
  \label{}
\begin{tabular}{@{\extracolsep{5pt}}lc}
\\[-1.8ex]\hline
\hline \\[-1.8ex]
 & \multicolumn{1}{c}{\textit{Dependent variable:}} \\
\cline{2-2}
\\[-1.8ex] & Citation.Counts.on.Google.Scholar \\
\hline \\[-1.8ex]
 Year.Distance.from.2020 & 4.769$^{***}$ \\
  & (0.654) \\
  & \\
 ConferenceSciVis & $-$69.884$^{***}$ \\
  & (9.983) \\
  & \\
 ConferenceVAST & $-$28.003$^{***}$ \\
  & (8.698) \\
  & \\
 ConferenceVis & $-$64.277$^{***}$ \\
  & (7.601) \\
  & \\
 PaperTypeJ & 25.274$^{***}$ \\
  & (8.605) \\
  & \\
 Number.of.Authors & $-$0.751 \\
  & (1.563) \\
  & \\
 Cross.type.CollaborationTrue & $-$4.143 \\
  & (6.418) \\
  & \\
 Cross.country.CollaborationTrue & $-$5.817 \\
  & (6.765) \\
  & \\
 With.US.AuthorsTrue & 7.075 \\
  & (5.439) \\
  & \\
 AwardTrue & 51.561$^{***}$ \\
  & (11.857) \\
  & \\
 Constant & 52.831$^{***}$ \\
  & (14.493) \\
  & \\
\hline \\[-1.8ex]
Observations & 3,070 \\
R$^{2}$ & 0.063 \\
Adjusted R$^{2}$ & 0.060 \\
Residual Std. Error & 141.763 (df = 3059) \\
F Statistic & 20.517$^{***}$ (df = 10; 3059) \\
\hline
\hline \\[-1.8ex]
\textit{Note:}  & \multicolumn{1}{r}{$^{*}$p$<$0.1; $^{**}$p$<$0.05; $^{***}$p$<$0.01} \\
\end{tabular}
\end{table}

\begin{figure}
\centering
  \includegraphics[scale = 0.8]{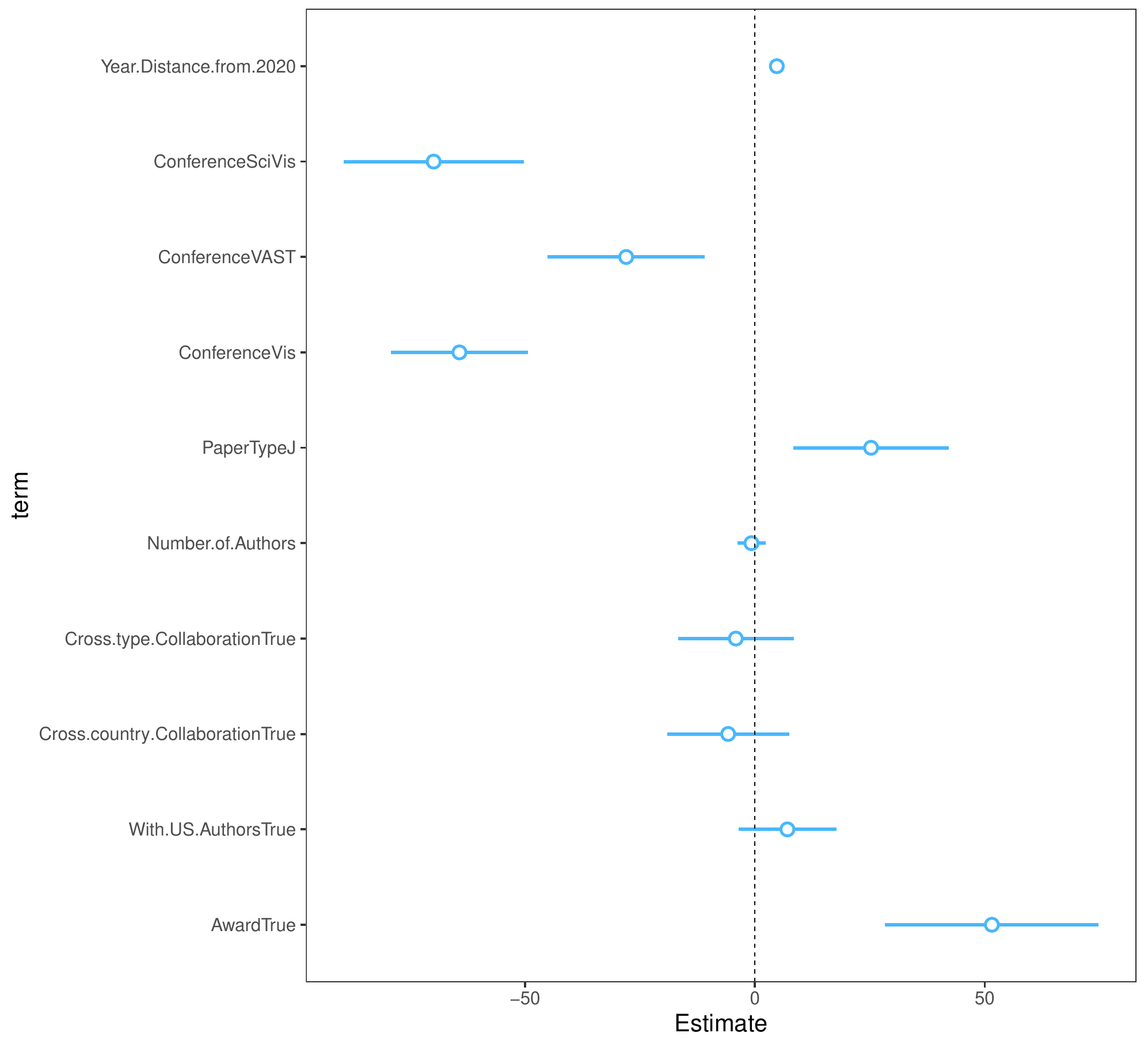}
    \caption{Regression results with Google Scholar citations. Estimates are unstandardized coefficients with 95\% confidence intervals.}
\label{fig:reg_openalex}
\end{figure}

\subsubsection{Log10 transformation on OpenAlex citations}

\begin{table}[!htbp] \centering
  \caption{Regression results with log10 transformation on OpenAlex citations. Estimates are unstandardized coefficients with standard error and $p$ values.}
  \label{}
\begin{tabular}{@{\extracolsep{5pt}}lc}
\\[-1.8ex]\hline
\hline \\[-1.8ex]
 & \multicolumn{1}{c}{\textit{Dependent variable:}} \\
\cline{2-2}
\\[-1.8ex] & citenum\_log10 \\
\hline \\[-1.8ex]
 Year.Distance.from.2020 & 0.015$^{***}$ \\
  & (0.002) \\
  & \\
 ConferenceSciVis & $-$0.442$^{***}$ \\
  & (0.035) \\
  & \\
 ConferenceVAST & $-$0.080$^{***}$ \\
  & (0.031) \\
  & \\
 ConferenceVis & $-$0.216$^{***}$ \\
  & (0.027) \\
  & \\
 PaperTypeJ & 0.190$^{***}$ \\
  & (0.031) \\
  & \\
 Number.of.Authors & $-$0.001 \\
  & (0.006) \\
  & \\
 Cross.type.CollaborationTrue & 0.016 \\
  & (0.023) \\
  & \\
 Cross.country.CollaborationTrue & 0.002 \\
  & (0.024) \\
  & \\
 With.US.AuthorsTrue & $-$0.014 \\
  & (0.019) \\
  & \\
 AwardTrue & 0.254$^{***}$ \\
  & (0.042) \\
  & \\
 Constant & 1.287$^{***}$ \\
  & (0.052) \\
  & \\
\hline \\[-1.8ex]
Observations & 3,070 \\
R$^{2}$ & 0.086 \\
Adjusted R$^{2}$ & 0.083 \\
Residual Std. Error & 0.504 (df = 3059) \\
F Statistic & 28.628$^{***}$ (df = 10; 3059) \\
\hline
\hline \\[-1.8ex]
\textit{Note:}  & \multicolumn{1}{r}{$^{*}$p$<$0.1; $^{**}$p$<$0.05; $^{***}$p$<$0.01} \\
\end{tabular}
\end{table}

\begin{figure}
\centering
  \includegraphics[scale=0.8]{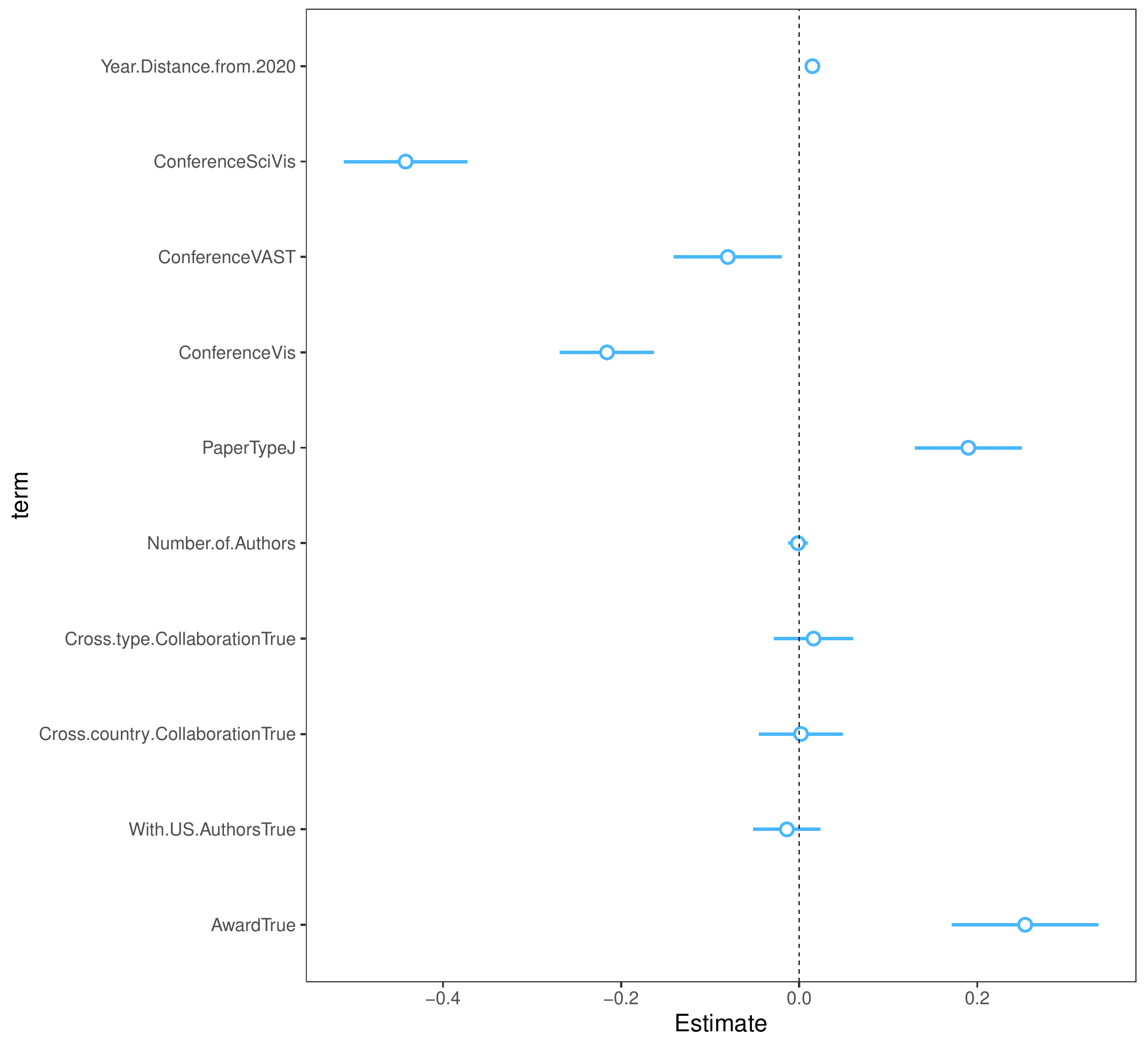}
    \caption{Regression results with log10 transformation on OpenAlex citations. Estimates are unstandardized coefficients with 95\% confidence intervals.}
\label{fig:reg_openalex}
\end{figure}




\printbibliography 